\documentclass[iop,twocolumn,tighten,numberedappendix,twocolappendix,revtex4]{emulateapj}
\usepackage{etoolbox}

\graphicspath{{./}{Plots/}}

\usepackage{subfigure}
\usepackage{graphicx}	
\usepackage{amsmath}	
\usepackage{amssymb}	

\newcommand{\software}[1]{\bgroup \vskip 6pt{\frenchspacing \font \foo=cmr10 \fontdimen2 \foo=3pt {\large \it Software: }#1\fontdimen2 \foo=3.33333pt}}

\usepackage{threeparttable}

\shorttitle{COSMOS-XS: Source catalog and number counts}
\shortauthors{D. van der Vlugt et al.}

\begin{document}

\title{An ultra-deep multi-band VLA survey of the faint radio sky (COSMOS-XS): Source catalog and number counts}

\author{D. van der Vlugt\altaffilmark{1}}
\altaffiltext{1}{Leiden Observatory, Leiden University, P.O. Box 9513, 2300 RA Leiden, the Netherlands}
\email{dvdvlugt@strw.leidenuniv.nl}

\author{H. S. B. Algera\altaffilmark{1}} 

\author{J. A. Hodge\altaffilmark{1}} 

\author{M. Novak\altaffilmark{2}}
\altaffiltext{2}{Max-Planck-Institut f\"ur Astronomie, K\"onigstuhl 17, D-69117 Heidelberg, Germany}

\author{J. F. Radcliffe\altaffilmark{3,4}}
\altaffiltext{3}{Jodrell Bank Centre for Astrophysics, The University of Manchester, SK11 9DL. United Kingdom}
\altaffiltext{4}{Department of Physics, University of Pretoria, Lynnwood Road, Hatfield, Pretoria 0083, South Africa}

\author{D. A. Riechers\altaffilmark{5,2}}
\altaffiltext{5}{Department of Astronomy, Cornell University, Space Sciences Building, Ithaca, NY 14853, USA}

\author{\\H. R\"{o}ttgering\altaffilmark{1}}

\author{V. Smol\v{c}i\'{c}\altaffilmark{6}}
\altaffiltext{6}{Department of Physics, University of Zagreb, Bijeni\^{c}ka cesta 32, 10002 Zagreb, Croatia}

\author{F. Walter\altaffilmark{2}}

\def\baselinestretch{1.0}

\begin{abstract}
We present ultra-deep, matched-resolution Karl G. Jansky Very Large Array (VLA) observations at 10 and $3 \, \rm{GHz}$ in the COSMOS field: the COSMOS-XS survey. The final 10 and $3 \, \rm{GHz}$ images cover $\sim \, 16 \, \rm{arcmin}^{2}$ and $\sim 180 \, \rm{arcmin}^{2}$ and reach median \textit{rms} values of $0.41 \, \mu \rm{Jy} \, \rm{beam}^{-1}$ and $0.53 \, \mu\rm{Jy} \, \rm{beam}^{-1}$, respectively. Both images have an angular resolution of $\sim \, 2.0^{\prime \prime}$. To fully account for the spectral shape and resolution variations across the broad bands, we image all data with a multi-scale, multi-frequency synthesis algorithm. We present source catalogs for the 10 and $3 \, \rm{GHz}$ image with 91 and 1498 sources, respectively, above a peak brightness threshold of $5 \sigma$. We present source counts with completeness corrections included that are computed via Monte Carlo simulations. 
Our corrected radio counts at $3 \, \rm{GHz}$ with direct detections down to $\sim \, 2.8 \, \mu$Jy are consistent within the uncertainties with other results at 3 and 1.4 GHz, but extend to fainter flux densities than previous direct detections.
The ultra-faint $3 \, \rm{GHz}$ number counts are found to exceed the counts predicted by the semi-empirical radio sky simulations developed in the framework of the SKA Simulated Skies project, consistent with previous P(D) analyses. Our measured source counts suggest a steeper luminosity function evolution for these faint star-forming sources. The semi-empirical Tiered Radio Extragalactic Continuum Simulation (T-RECS) predicts this steeper evolution and is in better agreement with our results. The $10 \, \rm{GHz}$ radio number counts also agree with the counts predicted by the T-RECS simulation within the expected variations from cosmic variance. In summary, the multi-band, matched-resolution COSMOS-XS survey in the well-studied COSMOS field provides a high-resolution view of the ultra-faint radio sky that can help guide next generation radio facilities.
\end{abstract}




\section{Introduction}
\label{sec:Introduction}
Over the last decade, a number of multi-wavelength (UV to radio) studies have revealed that the star-formation history of the universe (SFHU, i.e., the total star-formation rate (SFR) per unit of co-moving volume) went through several phases. The SFR apparently rose after the first galaxies formed (e.g., \citealt{Bouwens_2014a, Bouwens_2015}) and reached its peak during the ``epoch of galaxy assembly" at $1 \, \lesssim \, z \, \lesssim \, 3$. Subsequently, the SFR density declined rapidly to the present (e.g., \citealt{Madau_2014} and references therein). It is well established that the majority of the star-formation activity happens in galaxies that lie on the main sequence (MS), exhibiting an intrinsic scatter of $\sim \, 0.3$ dex, between SFR and stellar mass (e.g., \citealt{Brinchmann_2004, Noeske_2007, Elbaz_2007}). 
Not only does the relation exist in the local universe (e.g., \citealt{Brinchmann_2004, Salim_2007}) but is found to hold to $z \, \sim \, 4$ or even highter, albeit with a strong redshift evolution (e.g., \citealt{Daddi_2007, Pannella_2009, Karim_2011, Speagle_2014, Schreiber_2015, Salmon_2015, Tomczak_2016, Kurczynski_2016}). An accurate measurement of the SFR and its relation to stellar mass at all epochs is key for a better understanding of galaxy evolution. 

Several tracers across the electromagnetic spectrum can be used to measure this SFR, each with their own unique strengths and weaknesses. For example, ultraviolet (UV) light originates mainly from massive stars and thus directly traces young stellar populations. UV based observations, however, require uncertain and significant model-dependent dust corrections to account for dust obscuration (e.g., \citealt{Carilli_2008, Siana_2008, Siana_2009, Magdis_2010, Chary_2010, Bouwens_2012}). Infrared (IR) observations trace the absorbed UV emission that is thermally reprocessed by dust surrounding newly formed stars. However, the resolution of IR observations is often insufficient to provide reliable multi-wavelength identifications. In addition, the presence of polycyclic aromatic hydrocarbon (PAH) emission features redshifted ($z \, > \, 0.8$) into the 24$\mu$m band, commonly used as an estimator for the total-IR emission, makes the required $k$-correction uncertain.

Long-wavelength radio emission is another potential tracer of recent star-formation (\citealt{Condon_1992}). Radio emission in galaxies below rest frame frequencies $\lesssim \, 30 \, \rm{GHz}$ is dominated by synchrotron radiation arising from cosmic ray electrons gyrating in the galaxy's magnetic field (e.g., \citealt{Sadler_1989, Condon_1992, Clemens_2008, Tabatabaei_2017}). These charged cosmic-ray particles are accelerated in shocks launched by supernovae of stars with M $> \, 8 \, M_{\odot}$ in SFGs. These massive stars have lifetimes of $\lesssim \, 3 \, \times \, 10^{\text{8}}$ years, so their supernova rates are proportional to the recent SFR. This is supported by the tight correlation observed in SFGs between IR emission, originating from dust that has been heated by young and massive stars, and radio emission (the IR-radio correlation; e.g. \citealt{deJong_1985, Helou_1985, Yun_2001, Bell_2003, Dumas_2011}). Deep radio observations in the synchrotron regime can thus be used to constrain SFRs. Radio observations can also provide high spatial resolution to allow for reliable counterpart matching. 
Distant galaxies in the GHz radio regime often have spectral energy distributions (SEDs) that can be parameterised by a featureless powerlaw, leading to a simple and robust $k$-correction.
Radio observations in the synchrotron regime thus offers a unique opportunity to study the star-formation history of the Universe (e.g., \citealt{Condon_2002, Seymour_2008, Smolcic_2009_a, Jarvis_2015, Rivera_2017, Novak_2017}) at a wavelength that is free from selection biases due to dust obscuration.

However, there are two challenges in using radio emission in the synchrotron regime as a tracer of star-formation. The first is the ``contamination'' by active galactic nuclei (AGN). It is hard to disentangle AGN and SFGs in the radio regime as an accreting supermassive black hole (SMBH) in an AGN can also accelerate the electrons that produce synchrotron emission. To attempt to correct for this, several methods have been developed for identifying different types of AGN and seperating them from SFGs (e.g., \citealt{Hickox_2009, Mendez_2013,Smolcic_2017b, Delvecchio_2017}; Algera et al. 2020) using mid-IR data, far-IR data, X-ray information and multi-band optical/IR photometry. Synchrotron emission can thus not only be used to study star-formation but also to study the blackhole accretion activity in the Universe (e.g., \citealt{Jarvis_2000, Smolcic_2009_b, Rigby_2011, McAlpine_2013, Best_2014, Delvecchio_2014, Delvecchio_2017, Sabater_2018}).

The second challenge is the depth achievable in radio surveys. With the improving capabilities of modern interferometers, along with sophisticated calibration techniques, surveys of the faint $\mu$Jy radio emitting objects are able to constrain the faint populations (e.g., \citealt{Rujopakarn_2016, Murphy_2017, Smolcic_2017, Owen_2018, Bondi_2018, Mauch_2019}).
At high flux densities, the source counts are well-constrained and found to be dominated by AGNs that follow a smooth power-law distribution down to $S_{1.4 \rm GHz} \,  \sim \,  1 \, \rm{mJy}$ (e.g., \citealt{Condon_1984, Windhorst_1990}). Below 1 mJy, the Euclidean normalised source counts flatten (e.g., \citealt{Richards_2000, Huynh_2005, Biggs_2006, Owen_2008, Bondi_2008, Padovani_2015}). It is now widely accepted that this flattening observed is due to the emergence of SFGs and radio-quiet AGN which begin to contribute significantly (e.g., \citealt{RowanRobinson_1993, Seymour_2004, Padovani_2009}). New deep radio observations and $P(D)$ analyses on confusion limited surveys show evidence of a further steepening of the number counts below $S_{1.4 \rm GHz} \sim 50 \, \mu$Jy (e.g., \citealt{Condon_2012, Vernstrom_2014, Vernstrom_2016, Smolcic_2017, Prandoni_2018, Mauch_2019}). The composition of this ultra faint radio population is still uncertain but it is expected from simulations and observations that the fraction of SFGs will become significant ($>$ 60 \%) below $S_{1.4 \rm GHz} \sim 100 \, \mu$Jy (\citealt{Smolcic_2017b}). 
Constraints on the ultra-faint radio populations at high resolution are useful for predictions for future radio surveys with new and upcoming facilities such the ASKAP, MeerKAT, ngVLA, and SKA, where source confusion noise may be an issue.

\begin{figure*}
\centering
\includegraphics[width=1.0\textwidth]{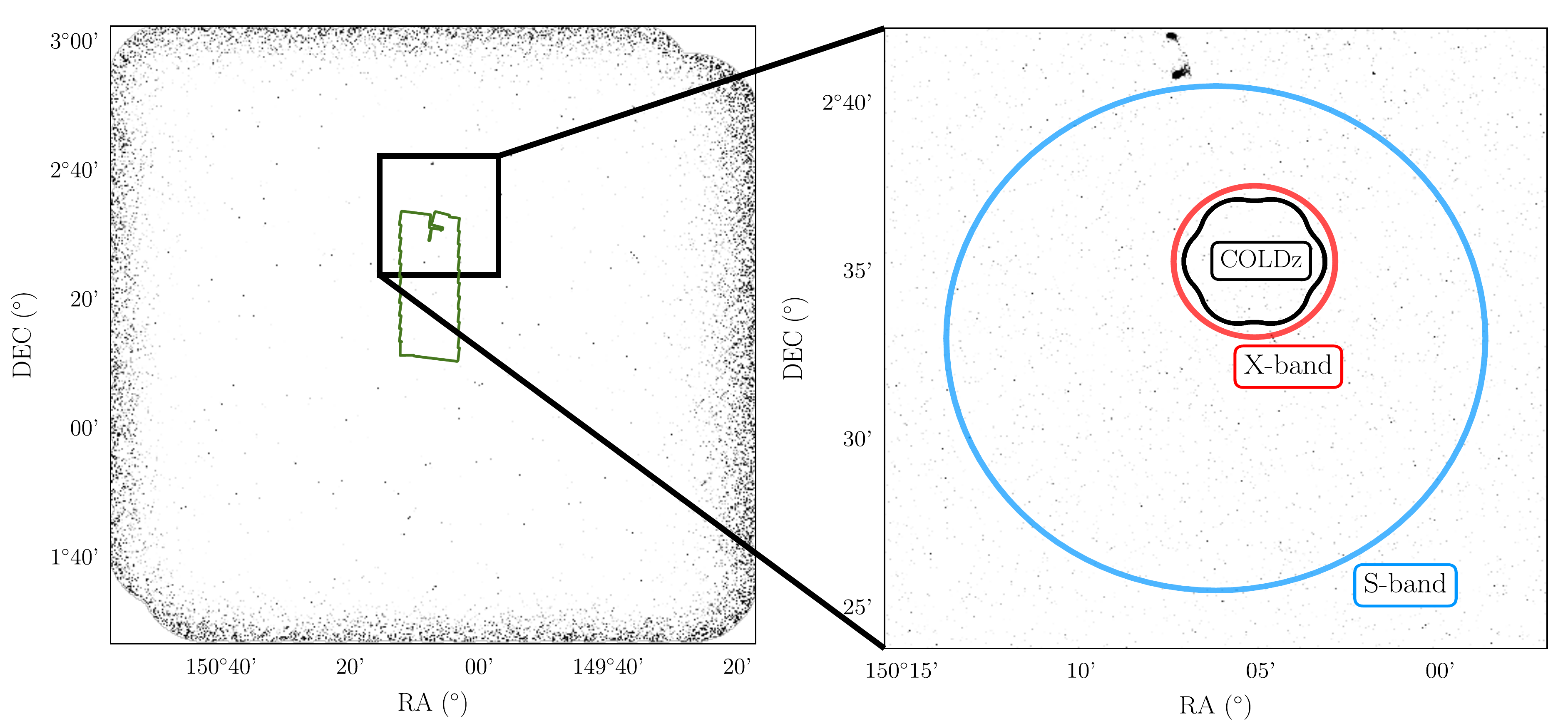}
\caption{The full $3 \, \rm{GHz}$ mosaic as imaged by \protect\cite{Smolcic_2017} of the 2 deg$^2$ COSMOS field. The green polygon indicates the footprint of the CANDELS WFC3 imaging (\protect\citealt{Nayyeri_2017}). The inset of $340 \, \rm{arcmin}^2$ shows the position of the X- and S-band primary beam. Additionally, the seven-pointing mosaic at $34 \, \rm{GHz}$, part of the COLDz project (\protect\citealt{Pavesi_2018}), is shown with the black contour.}
\label{fig:cosmos_field}
\end{figure*}

\begin{table*}
\centering
\caption{10 and $3 \, \rm{GHz}$ pointing centres and total observing time.}
\label{tab:point}
\begin{tabular*}{\textwidth}{@{\extracolsep{\fill}}lccccccr} 
	\hline
	Band & Central frequency & Configuration & \multicolumn{2}{c}{Centre (J2000)} & Total integration time & Primary beam FWHP & central r.m.s.\\
           & [GHz] &   &  RA & Dec  & [hours] & [arcmin] & [$\mu$Jy] \\
	\hline
S & 3 & B & 10$^{\rm h}$00$^{\rm m}$25$^{\rm s}$ & +02$^{\circ}$33$^{\prime}$00$^{\prime \prime}$ & 100 & 15 & 0.53 \\
X & 10 & C & 10$^{\rm h}$00$^{\rm m}$20$^{\rm s}$.7 & +02$^{\circ}$35$^{\prime}$17$^{\prime \prime}$ & 90 & 4.5  &  0.41\\

        \hline
\end{tabular*}
\end{table*}

Survey depth is also a challenge for radio surveys at observing frequencies $\geq \, 10 \, \rm{GHz}$. Such surveys measure flux densities closer to the rest-frame frequencies $\nu \, \geq \, 30 \, \rm{GHz}$ where the total radio emission is dominated by free-free radiation (e.g \citealt{Condon_1992, Murphy_2011, Klein_2018}) which constitutes the faintest part of the radio SED. Although more difficult to detect, free-free emission provides independent information on the star-formation process. Free-free emission directly originates from the H\textsc{II}-regions where massive stars form and thus provides a first-hand view on star-formation. In addition, unlike UV emission, dust obscuration plays only a minor role, and therefore free-free emission is potentially the most accurate tracer of star-formation (e.g. \citealt{Mezger_1967, Turner_1983, Turner_1985, Klein_1986, Kobulnicky_1999, Murphy_2012, Murphy_2015, Nikolic_2012}). 

An additional limitation of high frequency observations is the smaller primary beam area, as this area decreases with frequency ($\Omega_{\rm pb} \, \propto \, \nu^{-2}$). This limits the area that is covered at a certain depth. Low frequencies have therefore been favoured and radio continuum surveys probing free-free emission are still sparse in the literature. The majority of extragalactic radio surveys are instead conducted at 1.4 and $3 \, \rm{GHz}$, where the radio emission of galaxies is intrinsically brighter than at higher frequencies, and large areas can be imaged with a single pointing. In the last decade, several studies have been conducted to trace the synchrotron emission from galaxies (e.g., \citealt{Schinnerer_2007, Schinnerer_2010, Morrison_2010, Smolcic_2017, Prandoni_2018}). The COSMOS field has been observed with the Very Large Array (VLA) at $1.4 \, \rm{GHz}$ ($\sigma \, \sim \, 10 \-- 15 \, \mu\rm{Jy} \, \rm{beam}^{-1}$, \citealt{Schinnerer_2010}), and more recently at $3 \, \rm{GHz}$ with substantially better sensitivity ($\sigma \, \sim 2.3 \, \mu$Jy beam$^{-1}$), yielding about four times more radio sources compared to the $1.4 \, \rm{GHz}$ data (\citealt{Smolcic_2017}), see Fig. \ref{fig:cosmos_field}. Although these observations provide valuable data over the entire $2 \, \rm{deg}^2$ COSMOS field, enabling some of the most comprehensive studies to-date of the SFG and radio AGN populations, they are equivalent to $\sim \, 2$ hours per pointing. The resulting sensitivity only allows for the detection of typical SFGs out to $z \sim 1.5$ (\citealt{Novak_2017}), the epoch where the various SF tracers begin to diverge (e.g., \citealt{Ilbert_2013}). In order to observe MS galaxies over the full epoch of galaxy assembly ($z \, \sim \, 1 \-- 3$) a sub-$\mu$Jy survey is essential. 

The significantly increased bandwidths of the upgraded NSF's Karl G. Jansky Very Large Array (VLA) now allows for observations of the radio sky down to several hundred nJy beam$^{-1}$ sensitivities. 
We have taken advantage of these recent upgrades to the VLA to do an ultra-deep-matched-resolution survey in both the X- and S-band ($10 \, \rm{GHz}$ and $3 \, \rm{GHz}$). This COSMOS-XS survey is one of the deepest radio surveys to-date, reaching sub-$\mu$Jy sensitivities, and is $\sim \, 5$ times deeper than the previous $3 \, \rm{GHz}$ observations conducted in the COSMOS field (\citealt{Smolcic_2017}). When combined with the rich COSMOS multi-wavelength data, this survey thus yields a unique data set to test the composition of the faintest radio source populations that can currently be probed. 

The combined S- and X-band observations will enable us to study the properties and importance of AGNs and SFGs in a dust-unbiased way and to make predictions for the populations to be detected by future surveys. This paper (hereafter Paper I), describes the 10 and $3 \, \rm{GHz}$ observations and examines their implications for the ultra-faint source counts. In a companion paper (Algera et al. 2020; hereafter Paper II), we match the obtained catalogs with the multi-wavelength data available in the COSMOS field to distinguish between AGNs and SFGs. The obtained populations are then used to constrain the composition of the ultra-faint source counts. 

This paper is organised as follows: in Section \ref{sec:observations} we describe the VLA 10 and $3 \, \rm{GHz}$ observations, calibration, imaging and catalog extraction. In Section \ref{sec:cataloging}, we present the final images and describe the source-detection method and the compilation of the source catalog. This section also includes an analysis of the quality of the catalog. In Section \ref{sec:results} we discuss the derivation of the completeness-corrected radio source counts. Finally, Section \ref{sec:Conclusions} summarizes and concludes this work. Throughout this paper, the spectral index, $\alpha$, is defined as $S_{\nu} \, \propto \, \nu^{\alpha}$, where $S$ is the source flux density, and $\nu$ is the observing frequency. We assume a spectral index of $-0.7$ unless otherwise stated.

\section{Observations and data reduction}
\label{sec:observations}
\subsection{Observations}
The COSMOS-XS survey (see Table \ref{tab:point}) consists of the combination of a single deep S-band pointing centered on RA=10:00:25, Dec=+02$^{\circ}$33$^{\prime}$00$^{\prime \prime}$ (see Fig. \ref{fig:cosmos_field}) in the B-configuration and an additional single X-band pointing observed in the C-array centered on the COSMOS/AzTEC-3 protocluster at $z \, \sim \, 5.3$, coordinates RA=10:00:20.7 Dec=02$^{\circ}$35$^{\prime}$17$^{\prime \prime}$ (see Fig. \ref{fig:cosmos_field}). The chosen configurations provide a resolution of $\sim \, 2^{\prime \prime}$ in X- and S-band which is large enough to avoid resolving out faint sources. 
The X-band surveys overlaps with the COLDz survey (\citealt{Pavesi_2018, Riechers_2018}), one of the deepest $34 \, \rm{GHz}$ continuum surveys to date, covering $\sim \, 10 \, \rm{arcmin}^{2}$. The S-band pointing is chosen to be slightly offset from the X-band pointing to overlap more with the CANDELS-COSMOS field (\citealt{Nayyeri_2017}).
The X- and S-bands were observed for 90h and 100h, respectively. These data were taken between 4 Dec 2014 and 27 Feb 2016 with the individual observations constituting either two- or five-hour observing blocks. For the X-band observations, J1024-0052 was used as the phase calibrator, while for the S-band observations, J0925+0019 was used as the phase calibrator. For both bands, 3C286 served as both the flux and bandpass calibrator. The X-band covers a bandwidth of $4096 \, \rm{MHz}$ centered at $10 \, \rm{GHz}$, and is separated into 32 spectral windows $128 \, \rm{MHz}$ wide. All four polarization products were recorded and a 2s signal-averaging time was used. The S-band covers a bandwidth of $2048 \, \rm{MHz}$ centered at $3 \, \rm{GHz}$, and is separated into 16 spectral windows $128 \, \rm{MHz}$ wide. Again all four polarization products were recorded and a 5s signal-averaging time was used.

\subsection{Calibration}
\label{sec:calibration}
The calibration of the X-band and S-band visibilities was performed using CASA\footnote{\url{http://casa.nrao.edu/}} version 5.0.0. Extensive use was made of the NRAO VLA reduction pipeline\footnote{\raggedright\url{https://science.nrao.edu/facilities/vla/data-processing/pipeline}}.
Radio frequency interference (RFI) constitutes the major uncertainty in our data calibration. We experimented with different methods and tools (\texttt{rflag}, \texttt{AOflagger}; \citealt{Offringa_2010}) to remove the RFI. However, the unmodified pipeline resulted in the best image. 

Before running the pipeline, Hanning smoothing was applied to lessen Gibbs ringing from strong spectral features such as strong, narrow RFI. The pipeline runs several flagging rounds to flag bad or unnecessary data such as the initial few integration points of a scan where not all antennas may be on source.
To flag RFI the pipeline performed several rounds of flagging using the \texttt{rflag} algorithm. The pipeline uses the \texttt{gencal} and \texttt{gaincal} tasks to derive the necessary calibration solutions. \texttt{setjy} was used to to set the flux scale of the observations (\citealt{Perley_2013}) and the flux density calibrator was 3C286. During the calibration, no time or bandwidth averaging was performed so as to minimize time and bandwidth smearing effects (see Section \ref{sec:bandwith_smearing}).

In the X-band two spectral windows (SPWs), SPW 32 and SPW 33, were heavily flagged ($>$ 85\%) because of RFI. For the S-band only SPW 4 was found to be heavily corrupted by RFI and was flagged almost entirely by the pipeline. After the pipeline was run on the 10 and $3 \, \rm{GHz}$ data, the target field was split off using the task \texttt{split} and, respectively, 24.9\% and 25.8\% of the split off data was flagged.

\subsection{Bandwidth and time smearing}
\label{sec:bandwith_smearing}
An antenna receiver has a finite bandwidth which causes bandwidth smearing. This effect radially smears peak brightness while the integrated flux densities are conserved. Bandwidth smearing is a function of distance from the pointing center. The theoretical prediction from \cite{Condon_1998} for the reduction of peak response is given by $I/I_{0} \, = \,  1/\sqrt{1+0.46\beta^{2}}$, where $\beta \, = \,  (\Delta \nu / \nu_{0}) \, \times \, (\theta_{0}/\theta_{HPBW})$. If we calculate the reduction at 20\% of the peak primary beam sensitivity using the VLA channel width $\Delta \nu \, = \, 2 \, \rm{MHz}$, central frequency $\nu_{0} \, = \, 3 \, \rm{GHz}$ and a beam size of $\theta_{HPBW} \, = \,  2^{\prime \prime}$, we find an offset of $\sim \, 1$\%. When we compare the peak brightness over the total flux density for point-like sources ($0.9 \, \leq \, S_{\rm peak}/S_{\rm int}$) with $\rm{S/N} \, > \, 6$ as a function of distance from the pointing center, we find that an offset of $\sim \, 4$\% is present. This is, however, not distance dependent, and thus unlikely to be related to bandwidth smearing.

For the $10 \, \rm{GHz}$ observations, we calculate a reduction of $\sim \, 0.01$\% at 20\% of the peak primary beam sensitivity using the VLA channel width $\Delta \nu \, = \, 2 \, \rm{MHz}$, central frequency $\nu_{0} \, = \, 10 \, \rm{GHz}$ and a beam size of $\theta_{HPBW} \, = \, 2^{\prime \prime}$. The $S_{\rm peak}/S_{\rm int}$ distribution of point-like sources ($0.9 \, \leq \, S_{\rm peak}/S_{\rm int} $) with $\rm{S/N} \, > \, 6$ with distance also shows no distance dependent offset. Therefore, we do not apply any corrections at 10 and $3 \, \rm{GHz}$ for the bandwidth smearing effect.

The individual observations were concatenated after the calibration steps using the CASA task \texttt{concat}. These data were subsequently averaged in time using \texttt{split} with an averaging time of 21s for the X-band and 6s for the S-band. Averaging visibility data in the time domain causes a similar distortion to bandwidth smearing, but in the opposite direction (i.e. tangentially). The averaging time used should lead to an amplitude loss of at most one percent for a point source located at the first null of the primary beam due to loss of coherence, in the given VLA configurations.\footnote{\url{https://science.nrao.edu/facilities/vla/docs/manuals/}\newline\url{oss2013B/performance/fov/t-av-loss}}

\subsection{Imaging}
\begin{table}
\centering
\caption{Overview of the wide-field imaging parameters for the 10 and $3 \, \rm{GHz}$ image.}
\label{tab:imaging}
\begin{tabular}{lcccr} 
	\hline
	Band & robust & pixel size & \textit{w}-planes & restoring beam \\
           &   & [arcsec]  &   &  [arcsec] \\
	\hline
S & 0.5  & 0.4 & 400 & 2.14 $\times$ 1.81\\
X & 0.5 & 0.42  & 128 &  2.33 $\times$ 2.01\\

        \hline
\end{tabular}
\end{table}
Imaging of the concatenated datasets was performed both with CASA task \texttt{clean} and standalone imaging algorithm \texttt{WSclean} (\citealt{Offringa_2014}). For our purposes, the main difference between the two algorithms is the method of taking into account the interferometric \textit{w}-term during deconvolution. We opted for \texttt{WSclean} to create the final images, based on its faster processing compared to CASA's \texttt{clean}. However, differences between the CASA \texttt{clean} and \texttt{WSclean} images are minimal \citep[see also][]{Offringa_2014}, hence the choice of algorithm has no effect on the end product.

\texttt{WSclean} produces images by jointly gridding and deconvolving the measurement set, which is called joined-channel deconvolution (\citealt{Offringa_2017}). Spectral behavior of sources can be captured during deconvolution by setting the parameters {\fontfamily{qcr}\selectfont -channels-out} and {\fontfamily{qcr}\selectfont -join-channels}. The data is then imaged in separate channels across the band. During deconvolution, \texttt{WSclean} finds peaks in the full-band image and deconvolves these in each channel independently.

For the $10 \, \rm{GHz}$ image, we weight our image via Briggs weighting with a robust parameter of 0.5 (\citealt{Briggs_1995}) and apply \textit{w}-stacking using the minimum recommended number of 128 layers ({\fontfamily{qcr}\selectfont -nwlayers}).
We use the joined-channel deconvolution technique, specified by setting {\fontfamily{qcr}\selectfont -channels-out} to 32 (i.e. one per SPW) and specifying parameter {\fontfamily{qcr}\selectfont -join-channels}. A power law is fit to account for in-band spectral variations (similar to the multi-term multi-frequency synthesis algorithm used in CASA's \texttt{tclean} with nterms=2; \citealt{Rau_2011}). In addition, we utilize auto-masking of sources after first cleaning down to 3$\sigma$, whereupon masked sources are further cleaned down to 0.5$\sigma$ (parameters {\fontfamily{qcr}\selectfont -auto-mask} and {\fontfamily{qcr}\selectfont -auto-threshold}, respectively). The combination of these settings for {\fontfamily{qcr}\selectfont -auto-mask} and {\fontfamily{qcr}\selectfont -auto-threshold} is a good general setting for \texttt{WSclean}, and leaves almost no residuals. The resulting image reaches an r.m.s.-sensitivity of $\sigma \, = \, 0.41 \, \mu$Jy beam$^{-1}$ (Table \ref{tab:imaging}).

\begin{figure*}
\centering
\includegraphics[width=0.77\textwidth]{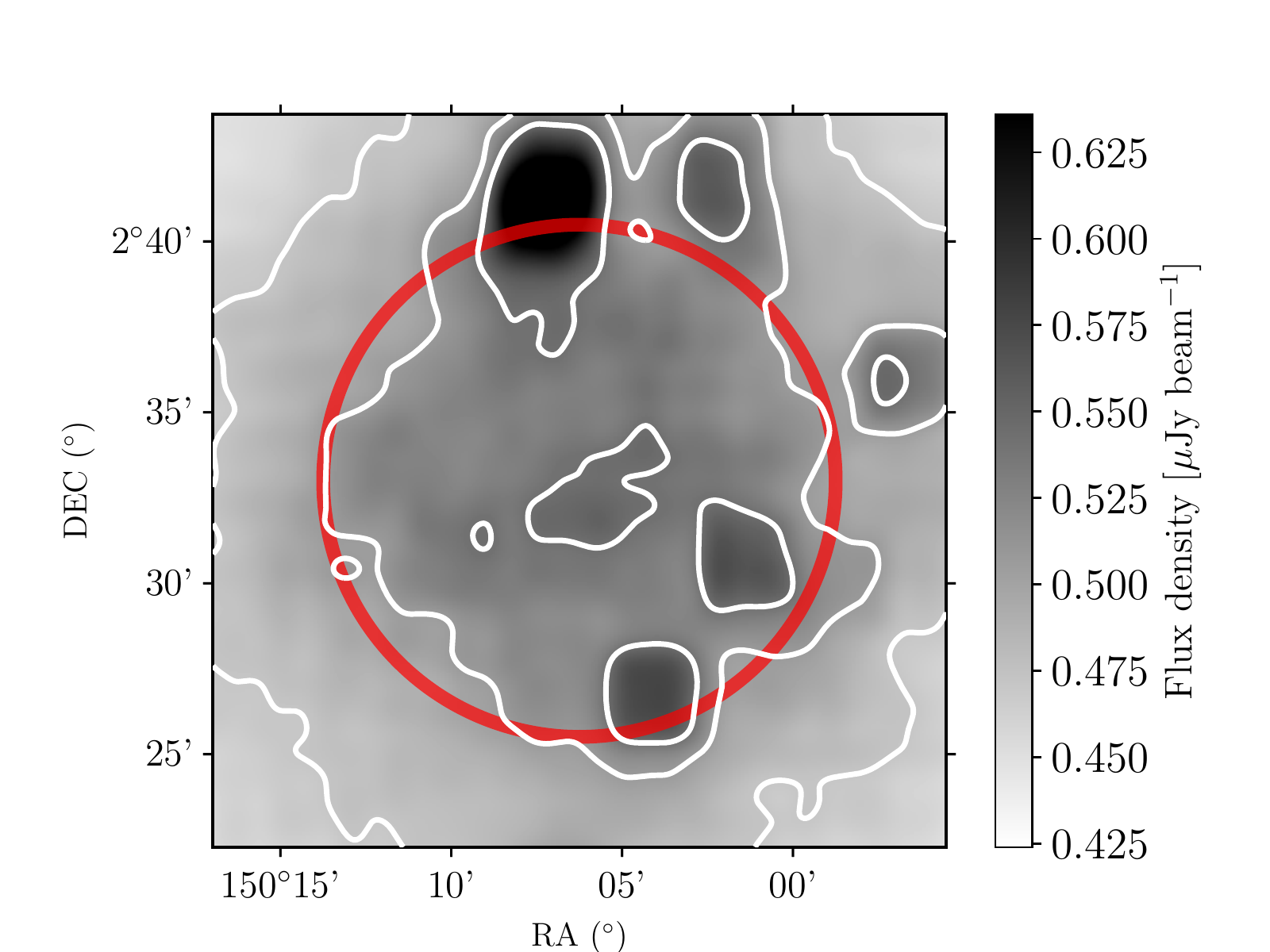}
\caption{The r.m.s. map of the $3 \, \rm{GHz}$ observation before primary beam correction. The image size is $460 \, \rm{arcmin}^2$. The r.m.s. map is created with PyBDSF. The red circle indicates the HPBW of the primary beam at $3 \, \rm{GHz}$, which corresponds to 15$^\prime$. The grey scale shows the r.m.s. noise from 0.8$\sigma$ to 1.2$\sigma$, where $\sigma \, = \, 0.53 \, \mu\rm{Jy} \, \rm{beam}^{-1}$. The contours are plotted at [0.47, 0.51, 0.54] $\mu$Jy beam$^{-1}$. 
}
\label{fig:Rms_S}
\end{figure*}

For the $3 \, \rm{GHz}$ image, we take a similar approach as described above, but with a few changes in the imaging parameters. The number of layers for \textit{w}-stacking is increased to 400. We find that \textit{w}-term artifacts persist for imaging with 128 layers and increasing the number to 400 solves these issues. We also do not fit the spectral variations with a powerlaw because of the large noise fluctuations at the beam edge caused by a bright source (see Fig. \ref{fig:Rms_S}). The {\fontfamily{qcr}\selectfont -channels-out} parameter is set to 16 (i.e. one per SPW). The resulting image reaches an r.m.s.-sensitivity of $\sigma \, = \, 0.53 \, \mu$Jy beam$^{-1}$ (Table \ref{tab:imaging}).

Because primary beam correction via projection-based gridding is not yet operational, we take the primary beam model from the CASA \texttt{widebandpbcor} task\footnote{This task was run by means of CASA version 5.3, which has an updated beam shape model.} which takes the frequency dependence of the beam into account.

\begin{figure*}
\centering
\includegraphics[width=0.77\textwidth]{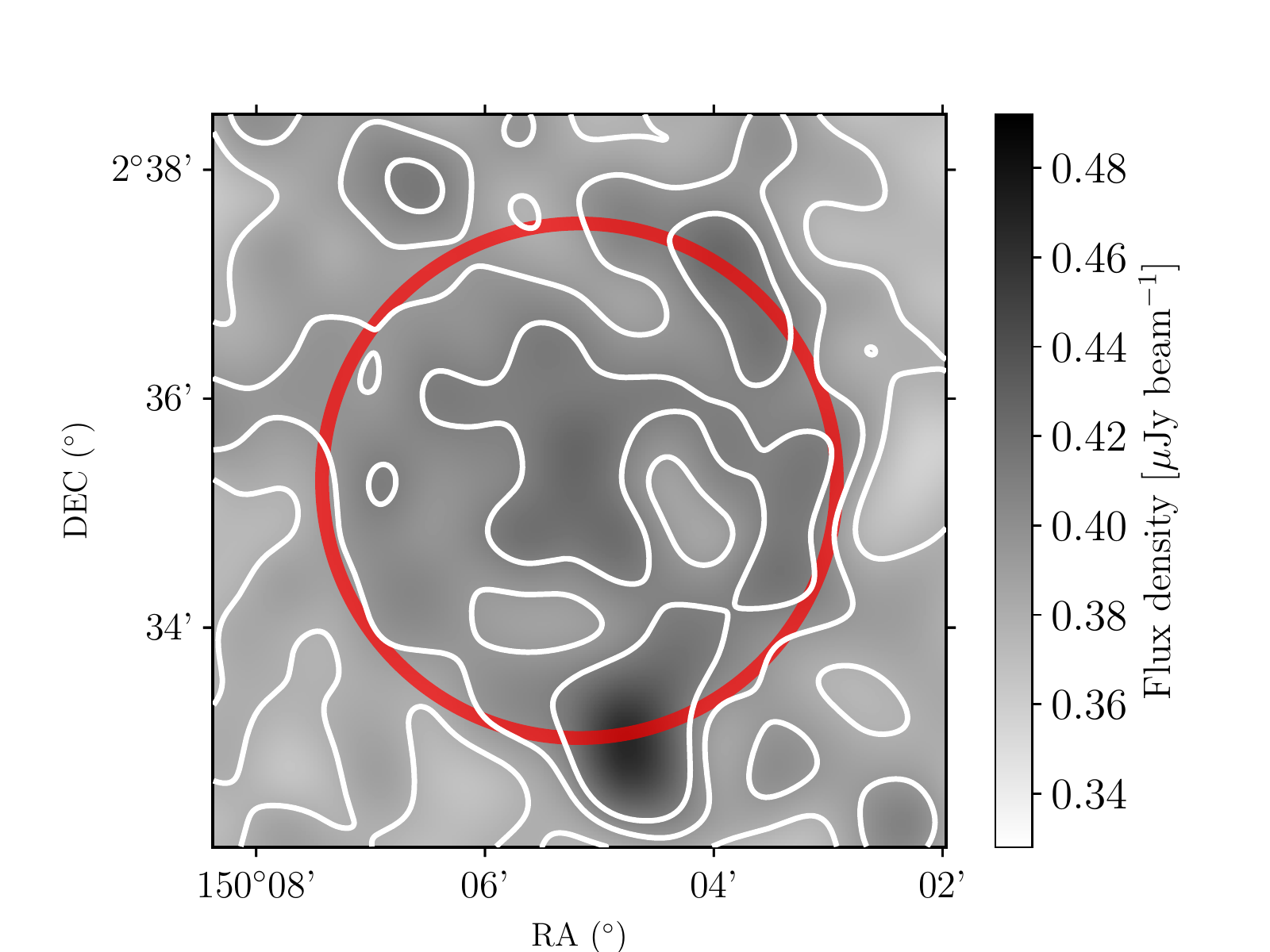}
\caption{The r.m.s. map of the $10 \, \rm{GHz}$ observation before primary beam correction. The image size is $41 \, \rm{arcmin}^2$. The r.m.s. map is created with PyBDSF. The red circle indicates the HPBW of the primary beam at $10 \, \rm{GHz}$, which corresponds to $4.5^\prime$. The grey scale shows the r.m.s. noise from $0.8\sigma$ to $1.2\sigma$, where $\sigma \, = \, 0.41 \, \mu\rm{Jy} \, \rm{beam}^{-1}$ The contours are plotted at [0.38, 0.39, 0.41] $\mu \rm{Jy} \, \rm{beam}^{-1}$. 
}
\label{fig:Rms_X}
\end{figure*}

\begin{figure*}
\centering
\includegraphics[width=0.77\textwidth]{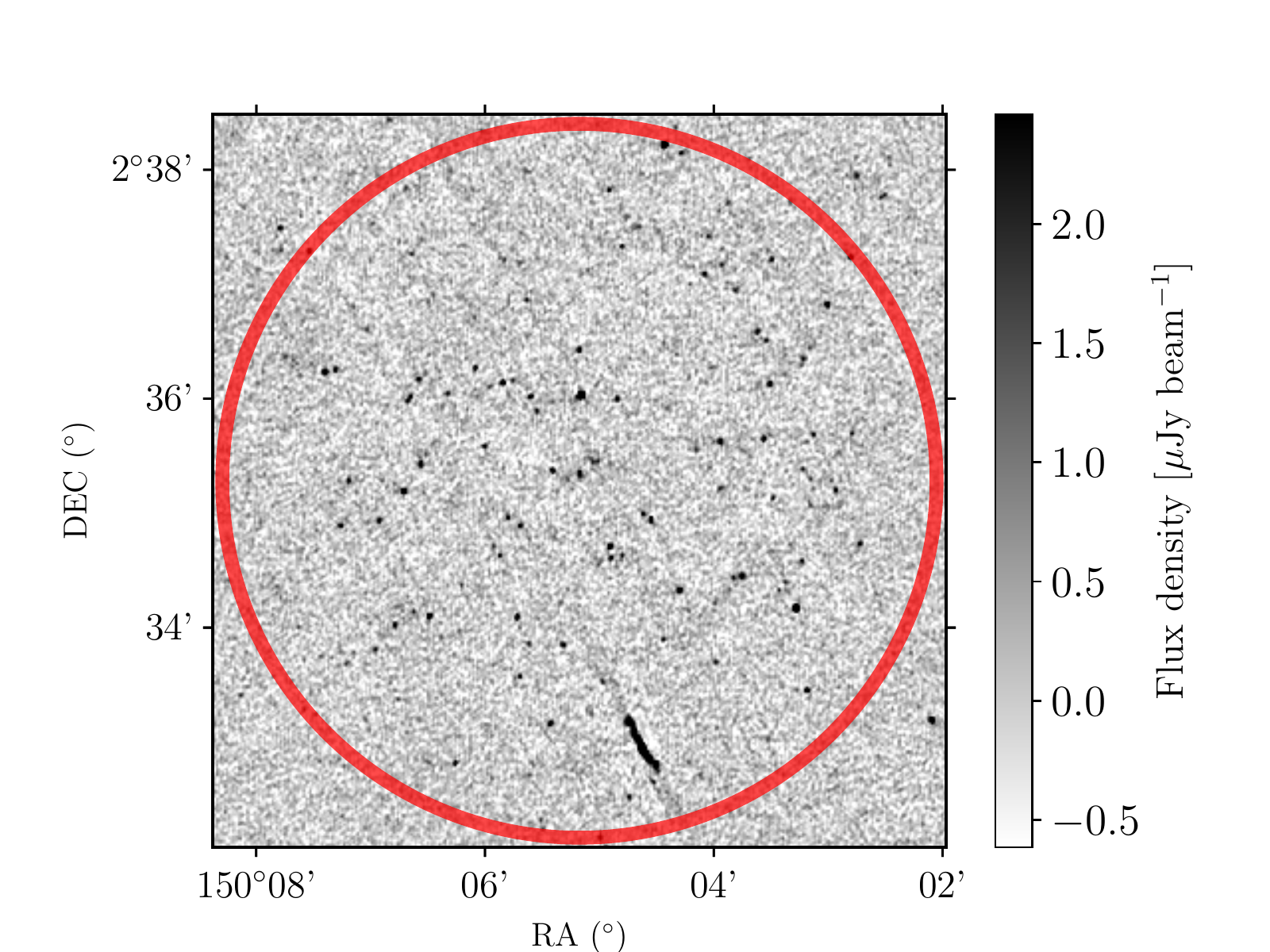}
\caption{Final calibrated $10 \, \rm{GHz}$ image before primary beam correction. The size is the same as in Fig. \ref{fig:Rms_X}. The red circle indicates the point where the primary beam sensitivity is 20\% of its peak. The grey-scale shows the flux density from $-1.5\sigma$ to $6\sigma$, where $\sigma \, = \, 0.41  \, \mu\rm{Jy} \, \rm{beam}^{-1}$ is the median r.m.s. within the primary beam FWHP. The corresponding brightness temperature r.m.s. value is $\sigma \, = \, $1.25 mK.
The image is matched in resolution and depth for a spectral index of $-0.7$ with the $3 \, \rm{GHz}$ image.
}
\label{fig:X_band_img}
\end{figure*}

\begin{figure*}
\centering
\includegraphics[width=0.77\textwidth]{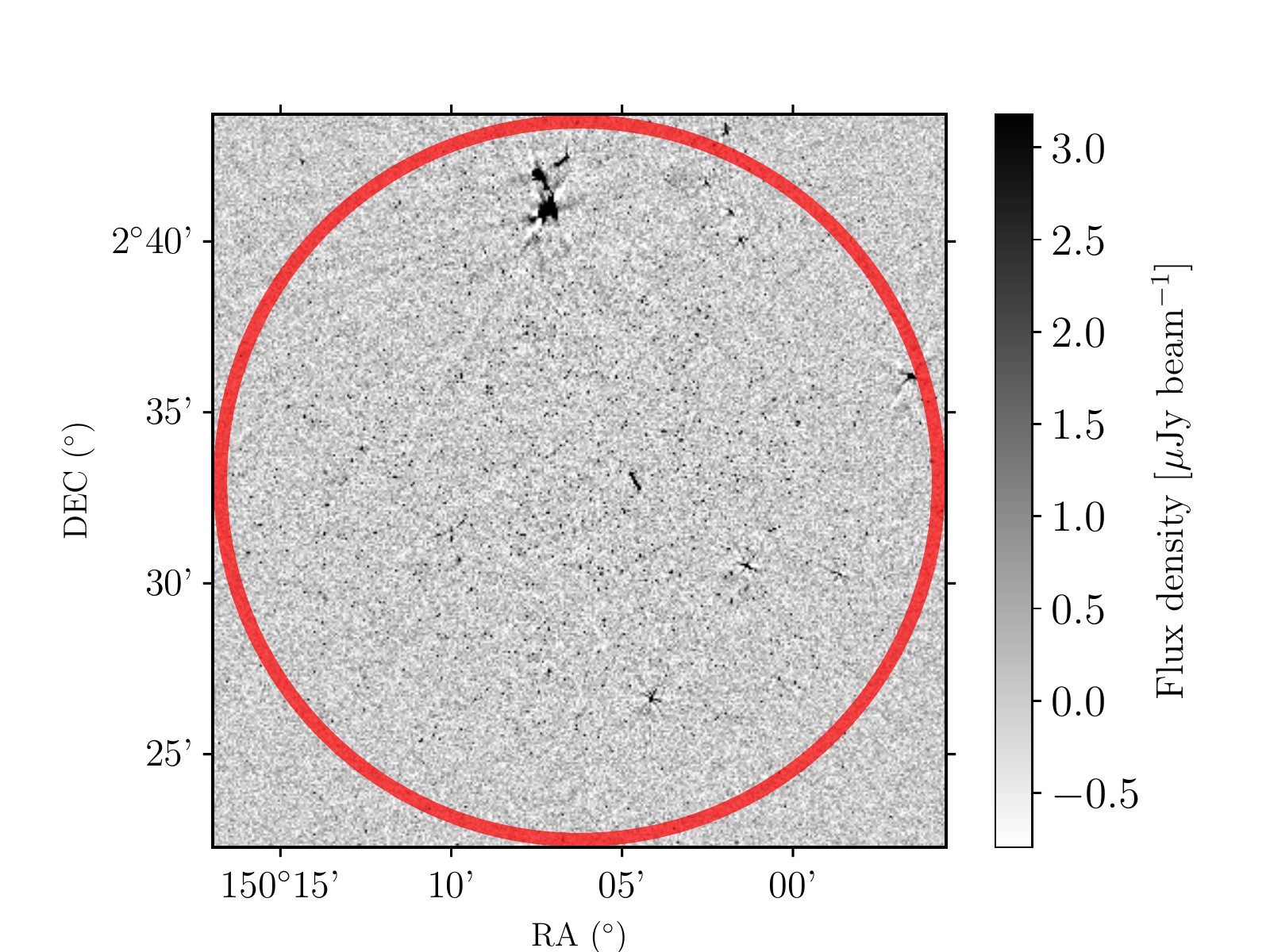}
\caption{Final calibrated $3 \, \rm{GHz}$ image before primary beam correction. The size is the same as in Fig. \ref{fig:Rms_S}. The red circle indicates the point where the primary beam sensitivity is 20\% of its peak. The grey-scale shows the flux density from $-1.5\sigma$ to $6\sigma$, where $\sigma \, = \,  0.53 \, \mu\rm{Jy} \, \rm{beam}^{-1}$ is the median r.m.s. within the primary beam FWHP. The corresponding brightness temperature r.m.s. value is $\sigma \, = \, $18.0 mK. 
}
\label{fig:S_band_img}
\end{figure*}

\subsection{Confusion}
Finally, we note that the source confusion is negligible in the deep $3 \, \rm{GHz}$ VLA observation (only the S-band image is considered as it has the highest source density). The beam size is $2\farcs14 \times 1\farcs81$, which results in $3.16 \, \times \, 10^6$ beams per deg$^{2}$. At the faintest flux density bin of our number count measurements for the S-band (see Sec. \ref{sec:source_counts}), we find $\sim \, 1 \, \times \, 10^4$ sources per deg$^{2}$. This translates to one source per $\sim \, 316$ beams, and implies that confusion is not an issue. Following \cite{Condon_2012}, source confusion becomes important at one source per 25 beams. This confusion limit depends on the slope of the scale-free power law approximation for the number counts
\begin{equation}
    \label{eq:number_counts}
    n(S) = k S^{-\gamma}\, ,
\end{equation}
\noindent where $K$ is the count normalization and $1 \, < \, \gamma \, < \, 3$ is the differential count slope (\citealt{Condon_2012}). For the calculation of the confusion limit we assumed a slope of $\gamma \, = \,  2.0$. We expect source confusion to contribute approximately $0.01 \, \mu \rm{Jy} \, \rm{beam}^{-1}$ to the noise following Equation (27) from \cite{Condon_2012}.

\section{Final image and Cataloging}
\label{sec:cataloging}
The final 10 and $3 \, \rm{GHz}$ images are shown in Fig. \ref{fig:X_band_img} and Fig. \ref{fig:S_band_img}, respectively. The central r.m.s. noise level is relatively smooth for both images (see also Fig. \ref{fig:Rms_X} and \ref{fig:Rms_S}). The $3 \, \rm{GHz}$ image shows a small number of artifacts (see e.g., the Northern part of the image shown in Fig. \ref{fig:S_band_img}). The artifacts are localized around bright sources and have little impact on the majority of the map, as can be seen from Fig. \ref{fig:Rms_S}.

\subsection{Source detection and characterization}
\label{sec:Source_detection_characterization}
We compiled a source catalog using PyBDSF\footnote{\url{http://www.astron.nl/citt/pybdsf/}} (\citealt{Mohan_2015}) to detect and characterize sources. We ran PyBDSF on the final image, using the pre-primary-beam-corrected image as the detection image.

PyBDSF identifies peaks of emission above a given threshold ({\fontfamily{qcr}\selectfont thresh\_pix}) that are surrounded by contiguous pixels with emission greater than a minimum ({\fontfamily{qcr}\selectfont thresh\_isl}). PyBDSF fits the identified island with one or more Gaussians, which are subsequently grouped into sources. This happens if all the pixels on the line joining their centers have a value greater than {\fontfamily{qcr}\selectfont thresh\_isl} and if the length of this line is less than half the sum of their FWHMs. The total flux density of the sources is estimated by adding those from the individual Gaussians, while the central position and source size are determined via moment analysis. 

The spatial variation of the image noise was estimated by sliding a box across the image in overlapping steps, calculating the root mean square (r.m.s.) of the pixels within the box and interpolating the values measured from each step. The resulting r.m.s. map provides PyBDSF with an estimate of the spatial variation of the image noise for detection thresholding purposes. The r.m.s. maps for the 10 (see Fig. \ref{fig:Rms_X}) and $3 \, \rm{GHz}$ image (see Fig. \ref{fig:Rms_S}) are determined with a sliding box of {\fontfamily{qcr}\selectfont rms\_box} = (120, 60) pixels and {\fontfamily{qcr}\selectfont rms\_box} = (400, 100) pixels (i.e. a box size of 400 pixels every 100 pixels), respectively. 

For source extraction, we used {\fontfamily{qcr}\selectfont thresh\_pix} = $5.0\sigma$ and {\fontfamily{qcr}\selectfont thresh\_isl} = $3.0\sigma$ (i.e. the limit at which flux density is included in the source for fitting). Fig. \ref{fig:Rms_X} and \ref{fig:Rms_S} illustrate the variation in r.m.s. noise determined across the image and shows the increase in local r.m.s. as a result of calibration artifacts near bright sources. We used the {\fontfamily{qcr}\selectfont group\_tol} parameter with a value of 1.0 as a larger value (allowing for larger sources to be fitted) was not necessary. Sources are classified as `S' for single sources and `M' for multiple Gaussian sources. The parameters of the Gaussian fitted to the source are reported by PyBDSF. 


For the $10 \, \rm{GHz}$ ($3 \, \rm{GHz}$) images, the total number of sources detected by PyBDSF within 20\% of the peak primary beam sensitivity is 93 (1498), of which 90 (1392) are single-component sources; sources fitted by a single Gaussian.

\subsection{Resolved sources}
\label{sec:Resolved_sources}
In order to determine whether our identified source components are resolved, we make use of the ratio between integrated flux density ($S_{\rm int}$) and peak brightness ($S_{\rm peak}$) which is a direct measure of the extension of a radio source. For an unresolved source, the peak brightness equals the total flux density. Noise influences the measurements of $S_{\rm int}$ and $S_{\rm peak}$ and, therefore, the $S_{\rm int}/S_{\rm peak}$ distribution gets broadened towards the low signal-to-noise end.
The effect of noise on the $S_{\rm int}/S_{\rm peak}$ ratio can be determined by performing Monte Carlo simulations in which simulated sources are added to the pre-primary-beam-corrected image and then retrieved the same way as the observed data. We simulate 100 mock sources and insert these in the real image. This is repeated 200 times, simulating 20000 sources in total. The method described below is used to derive the upper envelope of the $S_{\rm int}/S_{\rm peak}$ distribution for both the 10 and $3 \, \rm{GHz}$ image. 

Sources are injected as Gaussians with their FWHM equal to the beam size and they are thus unresolved by construction. We only insert point sources to quantify how the noise `resolves' unresolved sources. The peak brightness of the sources are drawn from the real source distribution to generate a realistic mock catalog. We fitted a power law of the form $n \, = \, a \, \times \, S^{-\gamma} \, + \, b$ (see Equation \ref{eq:number_counts}) to the binned measured peak brightness distribution, and draw fluxes between 3$\sigma$ and 60$\sigma$ randomly from this distribution. The mock sources are assigned a position that is at least 20 pixels ($\sim \, 8^{\prime \prime}$) away from both real sources and other mock sources. The position of the mock source also has to lie within the 20 per cent power point of the primary beam. The position was randomly chosen until both restrictions are satisfied. 

After all mock sources are inserted, we run PyBDSF with the exact parameters as performed for the real sources. Since the extraction is carried out on a map containing both real and mock emission, the real sources are always recovered and had to be filtered out in order to keep only simulated sources in the extracted catalog.

\begin{figure*}
     \begin{center}
        {%
            \label{fig:Bondi_X}
            \includegraphics[width=\columnwidth]{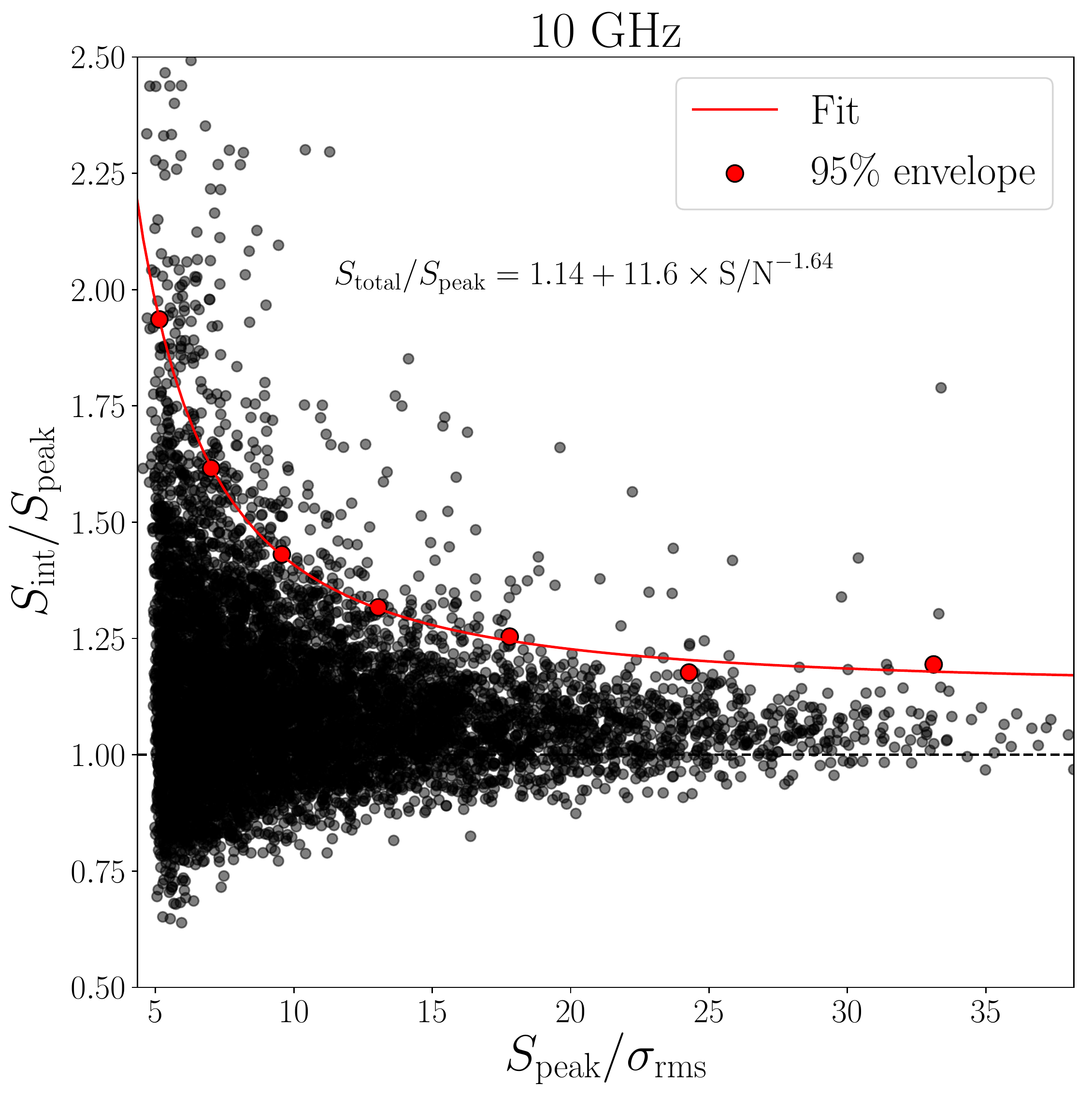}
        }%
        {%
           \label{fig:Bondi_S}
           \includegraphics[width=\columnwidth]{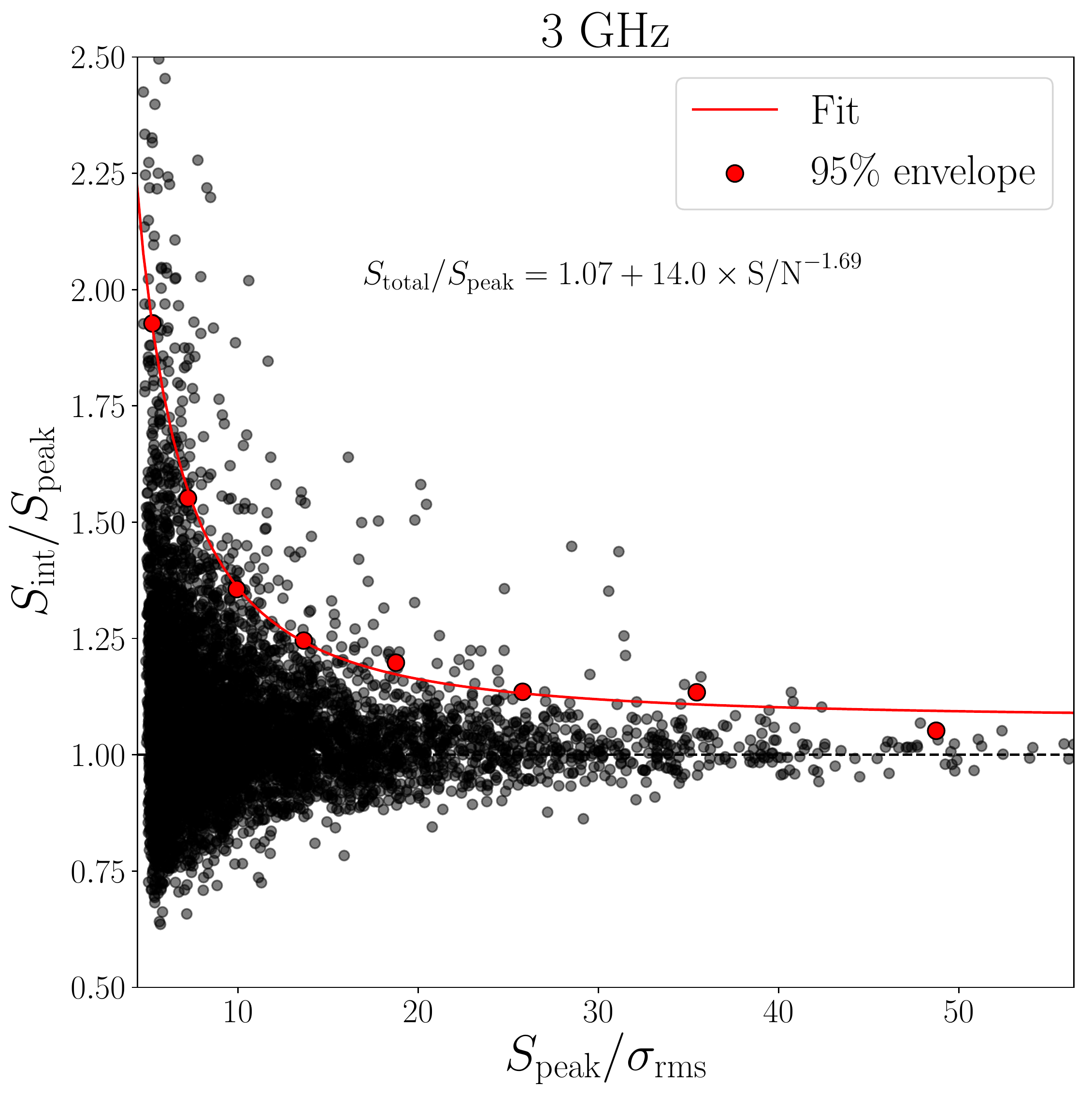}
        }
    \end{center}
    \caption{%
        The simulated ratio of integrated flux density to peak brightness as a function of signal-to-noise ratio for unresolved sources from the 200 Monte Carlo simulations for the $10 \, \rm{GHz}$ image (left) and the $3 \, \rm{GHz}$ image (right). For logarithmic bins in signal-to-noise ratio, the red points show the threshold below which 95 per cent of the sources lie in that bin. The red line shows the fit to this upper envelope. Applying these thresholds to the real data, we find that, respectively, 13\% and 32\% of the sources are resolved. 
     }%
   \label{fig:Bondi_X_S}
\end{figure*}

The recovered $S_{\rm int}/S_{\rm peak}$ distribution as a function of S/N is shown in Fig. \ref{fig:Bondi_X_S}. To determine the 95 per cent envelope in Fig. \ref{fig:Bondi_X_S}, a curve (red line) is fitted to the 95th percentile of logarithmic bins across the S/N, where $\text{N} \, = \, \sigma_{\rm local}$ as measured by PyBDSF. The shape of the envelope was chosen following \cite{Bondi_2008}. The fit for the $10 \, \rm{GHz}$ image simulations is given by $S_{\rm int}/S_{\rm peak} \, = \, 1.14 + 11.6 \, \times \, (\text{S/N})^{-1.64}$. The fit for the $3 \, \rm{GHz}$ image simulations is given by $S_{\rm int}/S_{\rm peak} \, = \, 1.07 + 14.0 \, \times \, (\text{S/N})^{-1.69}$. 

We consider sources from our catalog that lie above this envelope to be resolved. For the $10 \, \rm{GHz}$ image there are 12 (13\%) resolved sources and for the $3 \, \rm{GHz}$ image there are 475 (32\%) resolved sources. For sources that lie under the envelope, the integrated flux density was set equal to the peak brightness.
The resolved sources are flagged as resolved in the final catalog presented in Section \ref{sec:source_catalog}. Note that not all the PyBDSF sources with multiple Gaussian components are resolved by this criterion as each component is considered separately. Conversely, not all single-component sources are unresolved. 

\subsection{Flux boosting}
\label{sec:flux_boosting}
The noise fluctuations in the image may influence the flux measurements of the extracted sources. Since the counts of faint sources increase with decreasing flux density, there should be a sea of faint sources below the noise level that may influence the source extraction. There is therefore a probability that intrinsically faint sources are detected at higher flux because of noise fluctuations. This effect, called flux boosting, is extremely important at low S/N where flux measurement can be overestimated.

\begin{figure*}
     \begin{center}
        {%
            \label{fig:Fluxboost_X}
            \includegraphics[width=\columnwidth]{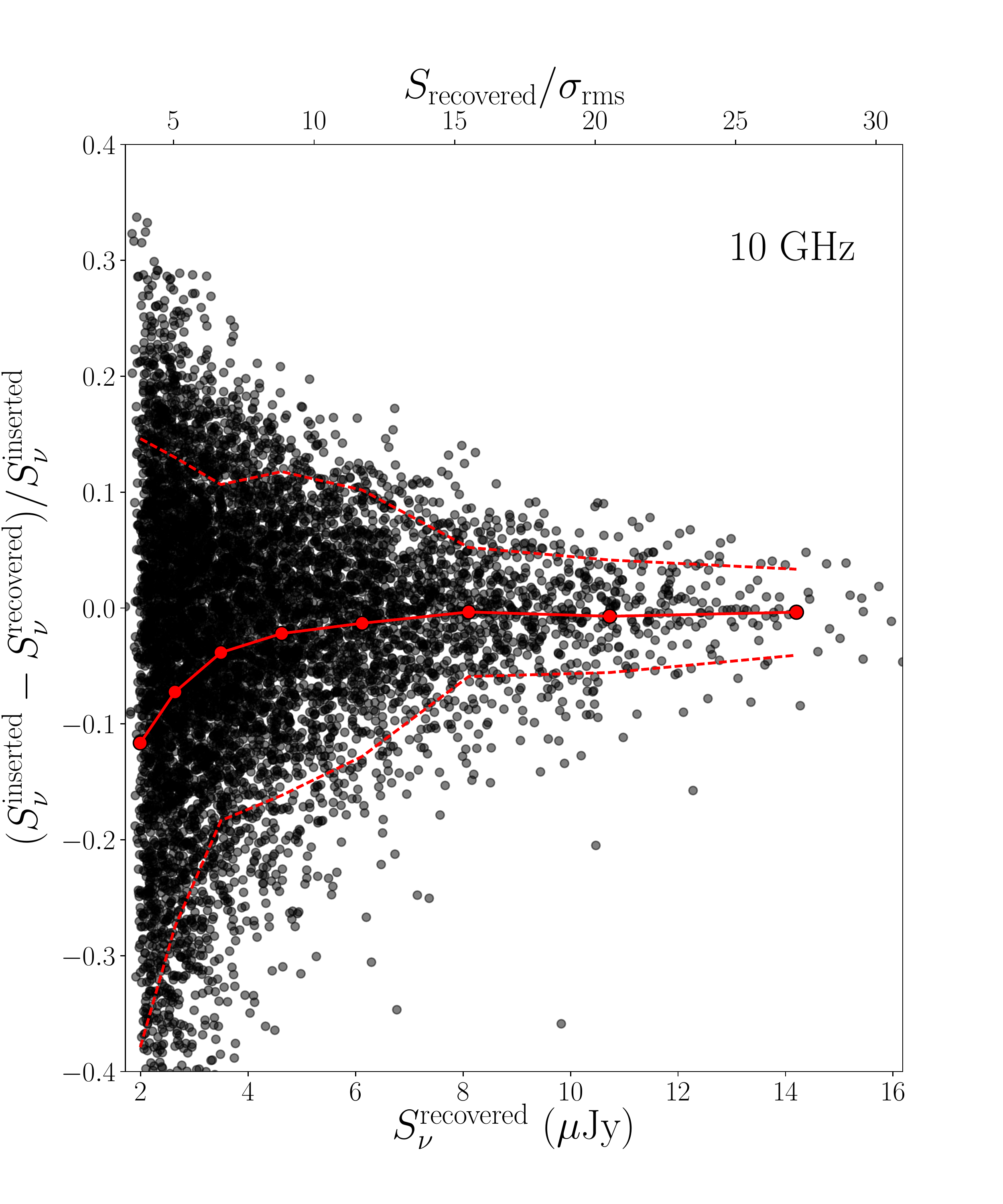}
        }%
        {%
           \label{fig:Fluxboost_S}
           \includegraphics[width=\columnwidth]{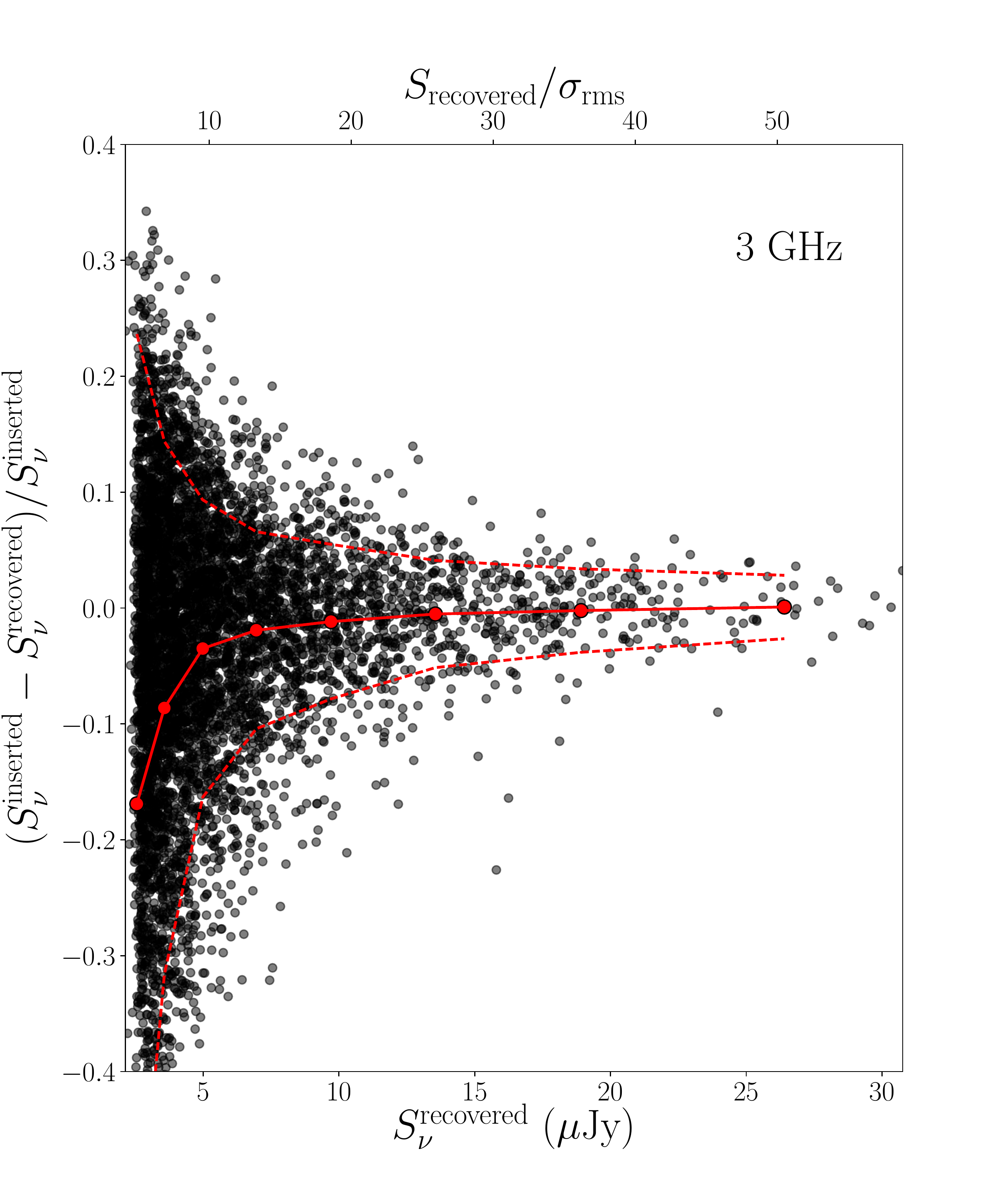}
        }
    \end{center}
    \caption{%
     The simulated ratio of (inserted flux density - recovered flux density) to inserted flux density as a function of the recovered flux density. The left panel shows the distribution for the $10 \, \rm{GHz}$ image and the right panel shows the distribution for the $3 \, \rm{GHz}$ image. The solid red line denotes the median of 8 logarithmic bins (indicated by the red points) across the flux density range and the dashed lines mark the 1$\sigma$ upper and lower bounds in those bins. The effect of flux boosting at the faint end is illustrated by the rapid downturn below about $3  \, \mu$Jy for $10 \, \rm{GHz}$ and $3 \, \rm{GHz}$. 
     }%
   \label{fig:Fluxboost_X_S}
\end{figure*}

The degree of flux boosting (the probability that faint sources are detected at higher flux density because of noise fluctuations) can be estimated by examining the output-to-input flux density of the simulations described in the above section. Fig. \ref{fig:Fluxboost_X_S} shows the distribution of the recovered flux density minus the inserted flux density normalized by the inserted flux density as a function of recovered flux density, where we calculated the mean and standard deviation in logarithmic bins. 

The effect for flux boosting is, as expected, greatest at the flux limit of our survey. For the $10 \, \rm{GHz}$ ($3 \, \rm{GHz}$) image, sources with $\rm{S/N} \, \simeq \, 5$ are boosted by 11\% (15\%) on average. The boosting effect quickly decreases with S/N, we find that sources with $\rm{S/N} \, \simeq \, 10$ are boosted by less then 5\% on average. Therefore, we do not correct for flux boosting. 

\begin{figure}
\centering
\includegraphics[width=1.0\columnwidth]{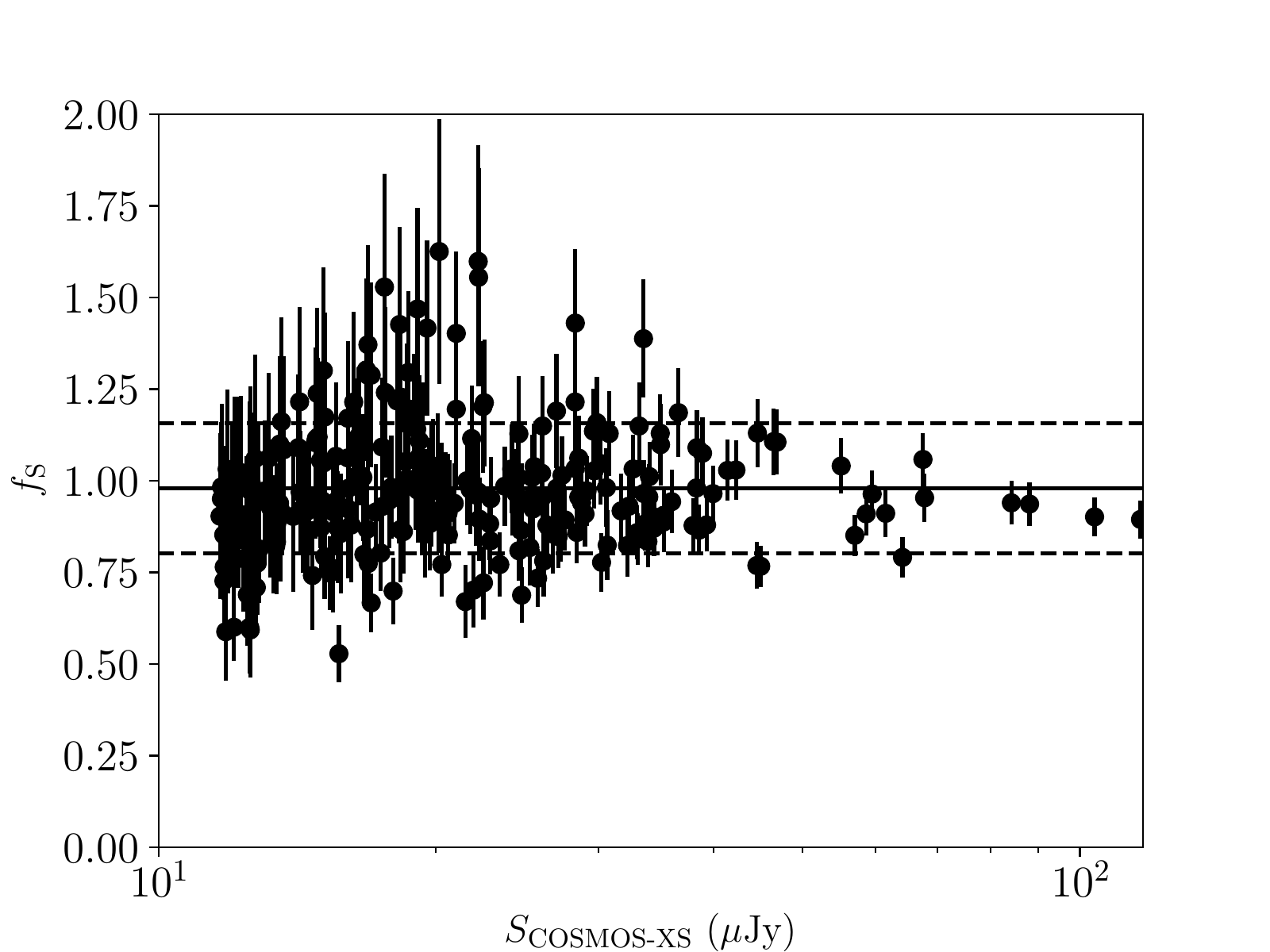}
\caption{Flux comparison between our sample and the VLA-COSMOS $3 \, \rm{GHz}$ Large Project (\protect\citealt{Smolcic_2017}). The median ratio of 0.98 with a standard deviation of 0.18 shows that the flux densities are on the same flux scale.}
\label{fig:S_band_fluxscale}
\end{figure}

\subsection{Flux density uncertainties at 3 GHz}
\label{sec:Flux_density}
In order to determine any systematic offsets, we have compared our flux densities to those of the VLA-COSMOS $3 \, \rm{GHz}$ Large Project. For the comparison, we selected only sources that could be detected at high signal to noise in the VLA-COSMOS $3 \, \rm{GHz}$ Large Project catalog which has a flux density limit of $S/\sigma > 5 = 11.5 \mu\text{Jy}$. We also consider only unresolved sources in the VLA-COSMOS $3 \, \rm{GHz}$ Large Project catalog to rule out resolution effects. This yielded a sample of 250 objects. For this subsample of sources, we determined the ratio of peak brightness between the COSMOS-XS survey and the VLA-COSMOS $3 \, \rm{GHz}$ Large Project $f_{\text{S}} = S_{\text{COSMOS-XS}}/S_{\text{Smolcic+2017}}$. The result of this comparison is shown in Fig. \ref{fig:S_band_fluxscale}. We measured a median ratio of 0.98 with a standard deviation of 0.18. The plot shows that the flux scale is in good agreement with the VLA-COSMOS $3 \, \rm{GHz}$ Large Project one over the entire flux range probed. We see also no systematic offsets in this subset with distance from the phase centre.

\subsection{Astrometry}
\begin{figure}
\centering
\includegraphics[width=\columnwidth]{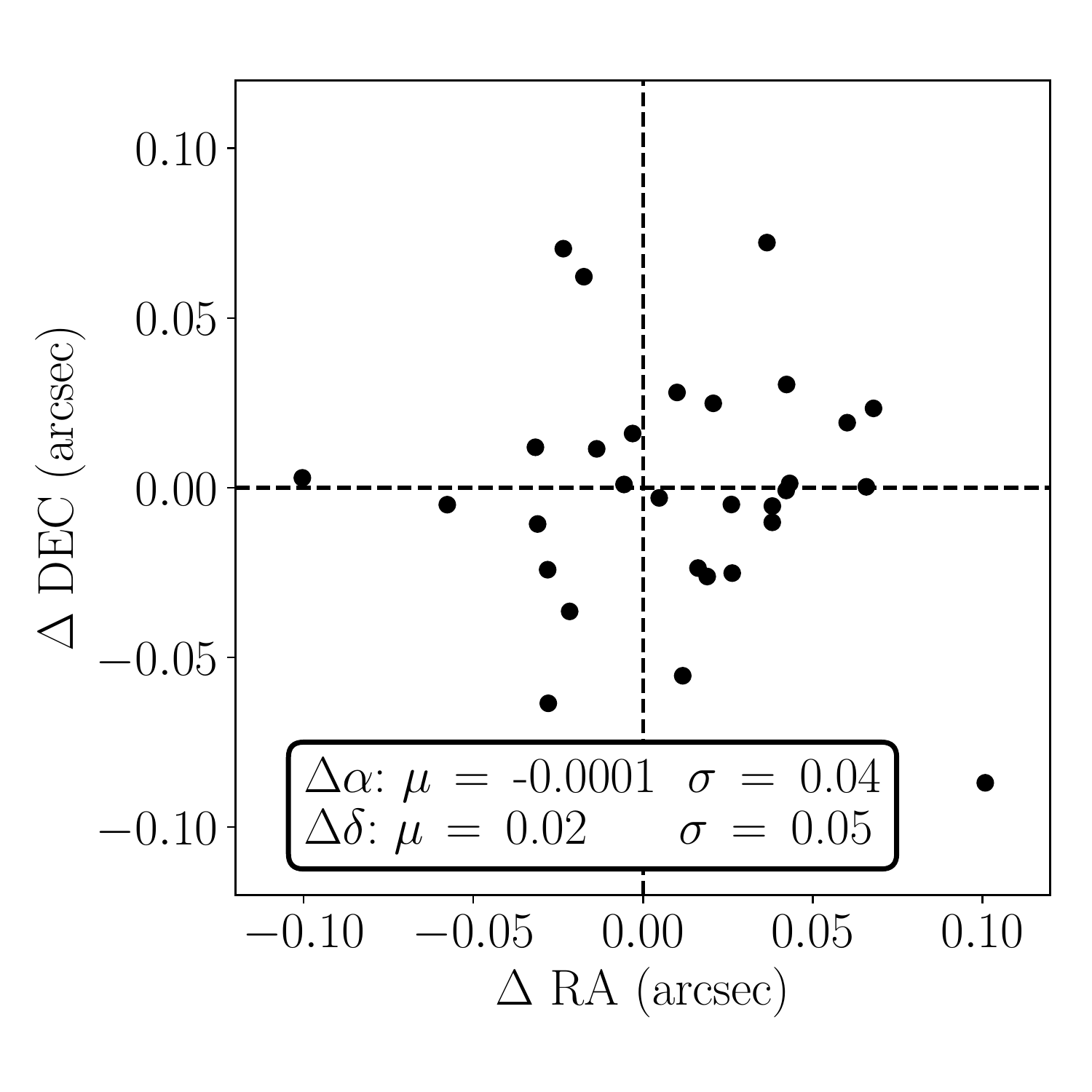}
\caption{Astrometry comparison between $3 \, \rm{GHz}$ and $1.4 \, \rm{GHz}$ VLBA data for 30 VLBA sources (\protect\citealt{Ruiz_2017}). Median and standard deviation for $\Delta$RA and $\Delta$Dec are
reported in the panel. The positions of sources detected at $3 \, \rm{GHz}$ are in excellent agreement with the posistions as reported in the Very Long Baseline Array (VLBA)- COSMOS $1.4 \, \rm{GHz}$ survey (\protect \citealt{Ruiz_2017}).}
\label{fig:ast_ra_dec}
\end{figure}

To assess our astrometric accuracy, we have compared the positions of 30 sources at $3 \, \rm{GHz}$ with $\rm{S/N} \, > \, 20$ with the positions detected in the Very Long Baseline Array (VLBA)- COSMOS $1.4 \, \rm{GHz}$ survey (\citealt{Ruiz_2017}). The matching radius used to find the $3 \, \rm{GHz}$ sources in the \cite{Ruiz_2017} catalog was $0.4^{\prime \prime}$. The results, shown in Fig. \ref{fig:ast_ra_dec}, yield an excellent agreement with a mean offset of $-0.0001^{\prime \prime}$ in $\Delta$RA and $0.02^{\prime \prime}$ in $\Delta$Dec. We find a standard deviation of $0.04^{\prime \prime}$ for $\Delta$RA and  $0.05^{\prime \prime}$ for $\Delta$Dec. We note that we did not correct the catalog entries for the offsets found.

\subsection{Completeness}
\label{sec:Completeness}
To quantify the completeness of the catalog, we performed another set of Monte Carlo simulations where we added simulated sources to the pre-primary-beam-corrected image. Injecting sources into the image allows us to account for the varying noise across the field. 
We use the same approach as in Section \ref{sec:Resolved_sources} but when the position of the source was determined, the input flux density of the source was reduced, dependent on position within the primary beam. By doing this we account for the decreasing sensitivity with increasing distance from the pointing centre due to the primary beam attenuation. 
Because the beam sensitivity is not uniform and decreasing, the effective area over which we are sensitive to a given flux density $S$ decreases rapidly with the flux density itself. By inserting primary beam corrected flux densities, the derived completeness correction automatically includes the effect of variation of sensitivity as a function of distance from the map centre. 

To allow for a better estimate of the completeness in terms of integrated flux densities, the mock sources also include a percentage of extended sources - Gaussians with FWHM larger than the beam size. The mock sources are therefore assigned, in addition to a peak flux density, a major axis, minor axis and an integrated flux density. To generate an angular size distribution for the mock sources, we draw randomly from two skewed Gaussians, one for the major axis and one for the minor axis. These Gaussians are determined by fitting to the normalized distribution of the fitted Gaussian parameters as measured for the real sources by PyBDSF.

The total flux density was chosen to be either their integrated flux density if resolved, or their peak brightness if unresolved. To determine whether a simulated source was resolved or unresolved, we use the same $S_{\rm int}/S_{\rm peak}$ envelope as described in Section \ref{sec:Resolved_sources}. 

Recovered mock sources are found by matching the retrieved catalog to the mock catalog. 
A retrieved source was considered to be matched with the inserted mock source if it was found within $0.5^{\prime \prime}$ of the inserted mock source position. Sources with a counterpart were flagged as recovered sources. 

\begin{figure*}
     \begin{center}
        {%
            \label{fig:Comp_X}
            \includegraphics[width=\columnwidth]{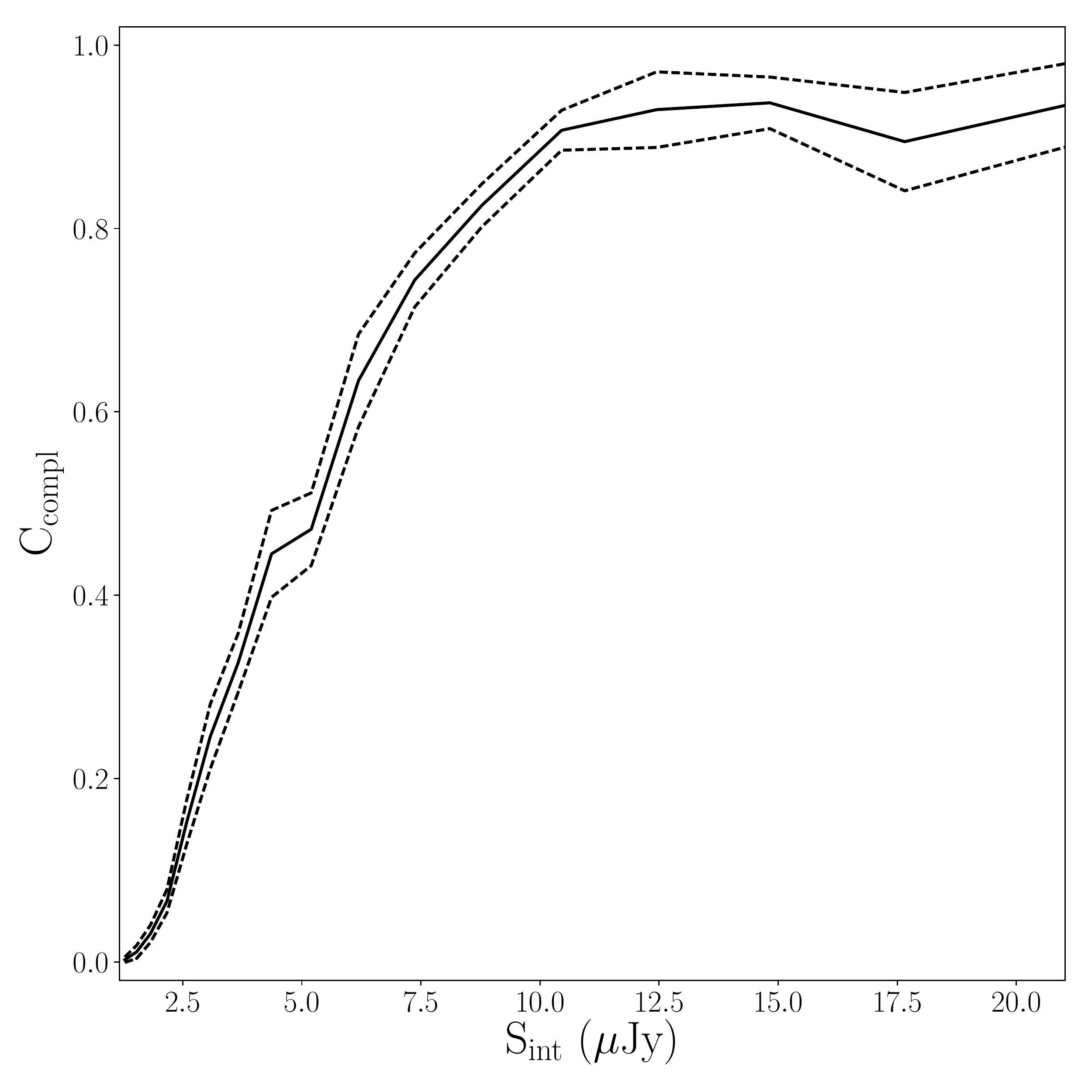}
        }%
        {%
           \label{fig:Comp_S}
           \includegraphics[width=\columnwidth]{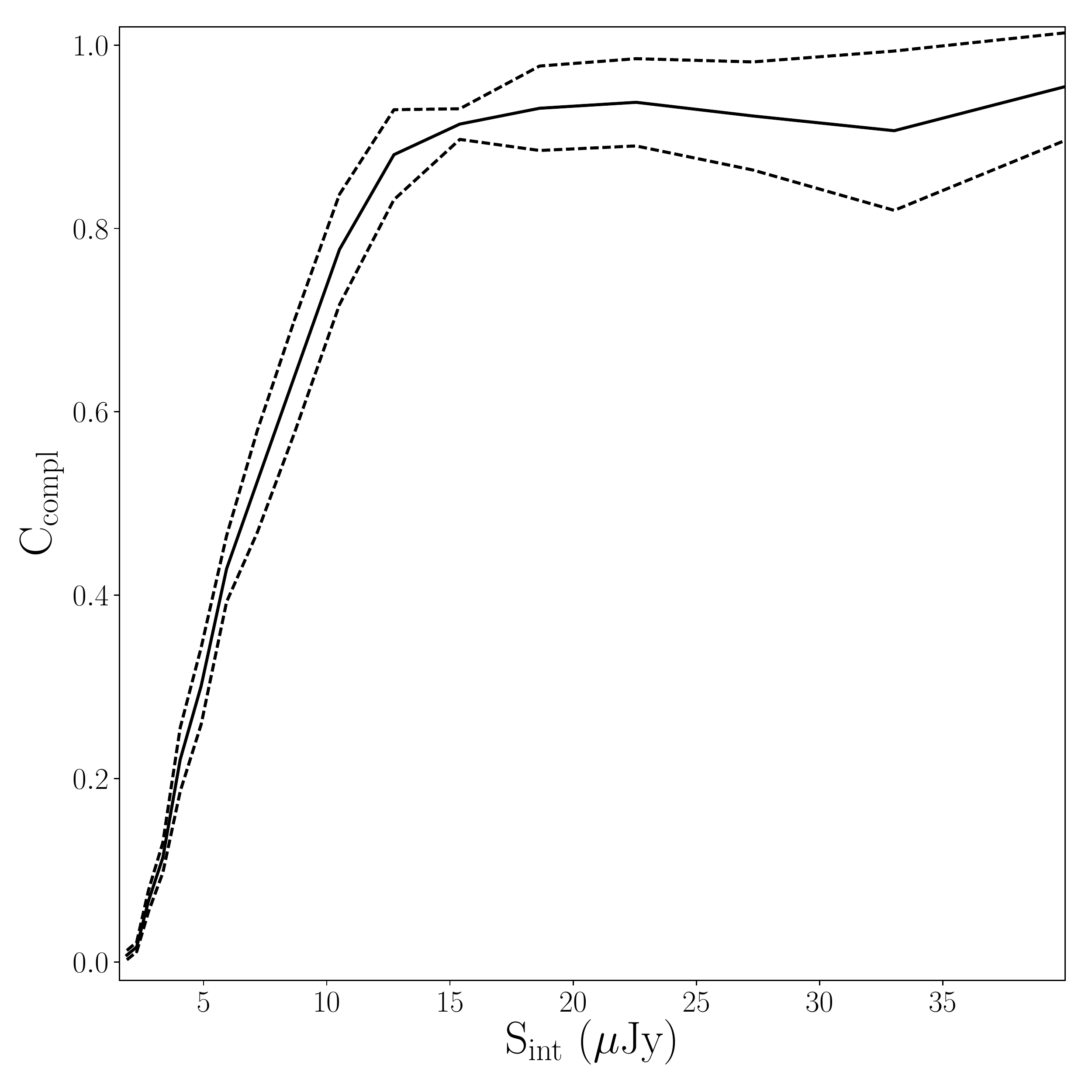}
        }
    \end{center}
    \caption{%
    Completeness of the $10 \, \rm{GHz}$ (left) and the $3 \, \rm{GHz}$ (right) source catalog as a function of flux density.
    The solid black line shows the mean completeness of all Monte Carlo runs and the dotted black lines show the standard deviation. The completeness is above $\sim \, 85$\% for $10  \, \mu$Jy and for $15 \, \mu$Jy for the $10 \, \rm{GHz}$ and the $3 \, \rm{GHz}$ source catalog, respectively. Note that the completeness does not reach $100$\% because effects of the primary beam are also included.
     }%
   \label{fig:Comp_X_S}
\end{figure*}

The completeness of a catalog represents the probability that all sources above a given flux density are detected. We have estimated this by giving the fraction of mock sources that are recovered using the same detection parameters. In Fig. \ref{fig:Comp_X_S} we plot the fraction of detected sources in our simulation as a function of integrated flux density, accounting for the primary beam. This detection fraction is largely driven by the variations in r.m.s. across the image and the primary beam. The error on the correction is the standard deviation of the calculated correction over all 200 realizations of the mock catalog. We thus estimate that the catalog is 85 per cent complete above a flux density of $10  \, \mu$Jy for the $10 \, \rm{GHz}$ image and above a flux density of $15  \, \mu$Jy for the $3 \, \rm{GHz}$ image.

\begin{figure}
\centering
\includegraphics[width=\columnwidth]{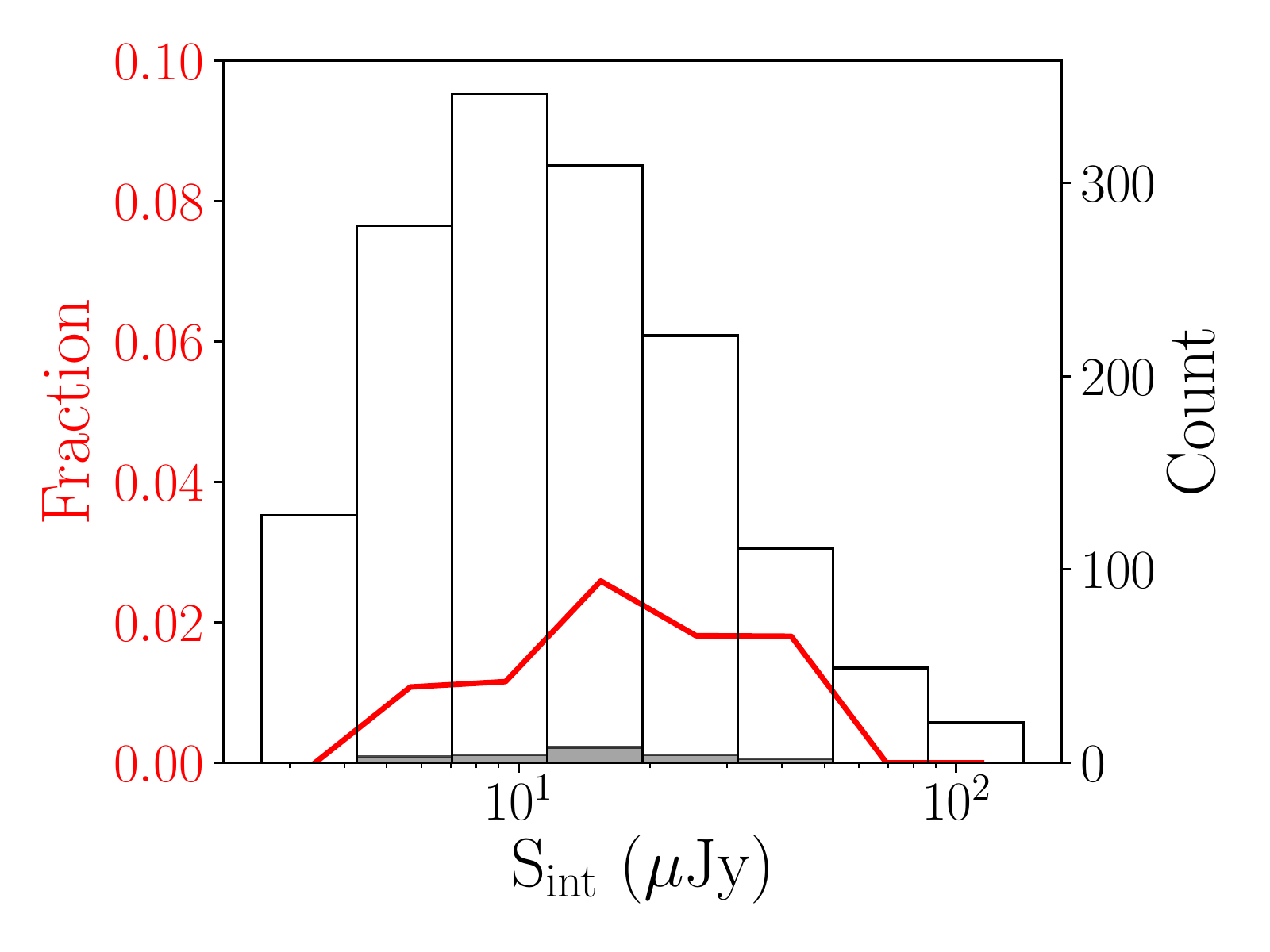}
\caption{Fraction of false detections (red line) as a function of flux density. The open (filled) histogram shows the number of components cataloged in the observed $3 \, \rm{GHz}$ map (detected in the inverted map). These data are also listed in Table \ref{tab:S_band_res}. The false detection rate is always smaller than 3\%, and is 1\% overall.}
\label{fig:False_dec_S}
\end{figure}

\subsection{Reliability}
\label{sec:Reliability}
The reliability of a source catalog indicates the probability that all sources above a given flux density are real sources and not accidental detections of background features or noise. To assess the false detection rate of our source extraction, we ran PyBDSF on the inverted (i.e., multiplied by $-$1) continuum map, within 20\% of the peak primary beam sensitivity, with the same settings used for the main catalog. Since there is no negative emission on the sky, every source detected in the inverted map is by definition a noise peak (i.e., a false detection). The false detection rate is determined from the number ratio of negative sources over positive sources per flux density bin. Errors are calculated based on Poissonian errors on the number of sources per flux density bin.

For the inverted $10 \, \rm{GHz}$ image, no sources were detected above our catalog threshold of 5$\sigma$, and thus the false detection rate was determined to be zero. The false detection rate found for the $3 \, \rm{GHz}$ image is shown in Fig. \ref{fig:False_dec_S} with the red line. The number counts for both the real and falsely detected sources are also shown. In total there are 22 negative detections for the $3 \, \rm{GHz}$ image, which results in a total false detection rate of $\sim \, 1$\%. These sources are located around bright sources and are thus likely caused by the artifacts surrounding these sources.  

\subsection{Source catalog}
\label{sec:source_catalog}
\begin{table*}
\centering
\caption{Sample of the $3 \, \rm{GHz}$ COSMOS-XS catalog.}
\label{tab:Catalog}
\begin{threeparttable}
\resizebox{1.0\textwidth}{!}{\begin{tabular}{p{0.35\columnwidth}p{0.14\columnwidth}p{0.12\columnwidth}p{0.12\columnwidth}p{0.12\columnwidth}p{0.22\columnwidth}p{0.22\columnwidth}p{0.12\columnwidth}p{0.14\columnwidth}p{0.14\columnwidth}}  
	\hline
	Source ID & RA & $\sigma_{\rm RA}$ & Dec & $\sigma_{\rm Dec}$ & S$_{\rm int}$ & S$_{\rm peak}$ & $\sigma_{\rm local}$ & Gaussians & Resolved \\
          & [deg] & [arcsec] & [deg] & [arcsec] & [$\mu$Jy] & [$\mu$Jy beam$^{-1}$] & [$\mu$Jy beam$^{-1}$] &  &  \\
         (1) & (2) & (3) & (4) & (5) & (6-7) & (8-9) & (10) & (11) & (12)\\
	\hline
    COSMOS-XS J100027.60+023634.21 & 150.11501 & 0.14907 & 2.6095 & 0.19592 & 4.04 $\pm$ 1.11 & 3.52 $\pm$ 0.67 & 0.65 & S & U \\
    COSMOS-XS J100014.35+023147.58 & 150.0598 & 0.03873 & 2.52988 & 0.04186 & 15.58 $\pm$ 1.03 & 13.74 $\pm$ 0.62 & 0.6 & S & U \\
    COSMOS-XS J100055.00+022849.80 & 150.22917 & 0.02603 & 2.4805 & 0.03019 & 52.54 $\pm$ 2.4 & 46.48 $\pm$ 1.44 & 1.41 & S & R \\
    COSMOS-XS J100031.26+023642.93 & 150.13026 & 0.03604 & 2.61192 & 0.04184 & 17.77 $\pm$ 1.16 & 16.04 $\pm$ 0.69 & 0.68 & S & U \\
    COSMOS-XS J100004.85+023559.51 & 150.02019 & 0.12507 & 2.59986 & 0.12439 & 7.01 $\pm$ 1.43 & 6.21 $\pm$ 0.85 & 0.84 & S & U \\
    COSMOS-XS J100056.66+022635.42 & 150.23607 & 0.02513 & 2.44317 & 0.03391 & 101.17 $\pm$ 3.56 & 76.85 $\pm$ 2.22 & 2.14 & S & R \\
    COSMOS-XS J100029.99+022903.64 & 150.12494 & 0.06062 & 2.48434 & 0.1729 & 5.85 $\pm$ 1.63 & 3.19 $\pm$ 0.66 & 0.66 & M & U \\
    COSMOS-XS J100046.61+023443.89 & 150.1942 & 0.2192 & 2.57886 & 0.10006 & 7.25 $\pm$ 1.37 & 5.54 $\pm$ 0.84 & 0.83 & S & U \\
    COSMOS-XS J100013.14+022550.66 & 150.05477 & 0.06633 & 2.43074 & 0.09338 & 22.27 $\pm$ 2.2 & 17.69 $\pm$ 1.34 & 1.32 & S & R \\
    COSMOS-XS J100004.59+023301.50 & 150.01914 & 0.05262 & 2.55042 & 0.06861 & 11.75 $\pm$ 1.31 & 11.05 $\pm$ 0.77 & 0.76 & S & U \\
        \hline
\end{tabular}}
\begin{tablenotes}
\item \textbf{Notes.} 
The format is the following: Column (1): Source name. Column (2) and (3): flux density-weighted right ascension (RA) and uncertainty. Column (4) and (5): flux density-weighted declination (Dec.) and uncertainty. Column (6) and (7): integrated source flux density and uncertainty, in $\mu$Jy. Column (8) and (9): peak brightness and uncertainty, in $\mu$Jy beam$^{-1}$. Column (10): the local r.m.s. noise. Column (11): number of Gaussian components. S refers to a single-Gaussian source. M refers to a a multi-Gaussian source. Column (12): a flag indicating the resolved parametrization of the source. `U' refers to unresolved sources and `R' to resolved sources.
\end{tablenotes}
\end{threeparttable}
\end{table*}
The final $10 \, \rm{GHz}$ catalog consists of 91 sources and the final $3 \, \rm{GHz}$ catalog consists of 1481 sources. Both catalogs are available as a table in FITS format as part of the online version of this article. Resolved sources are identified as described in Section \ref{sec:Resolved_sources}. Errors given in the catalog are the nominal fit errors reported by PyBDSM. A sample of the $3 \, \rm{GHz}$ catalog is shown in Table \ref{tab:Catalog}.

\section{Results}
\label{sec:results}
In this section, we report two results based on the produced catalogs: the 10 and $3 \, \rm{GHz}$ faint source counts. Further analysis of these data will be presented in future publications.

\subsection{Radio source counts}
\label{sec:Radio source counts}
We use the 10 and $3 \, \rm{GHz}$ catalogs to compute the source
counts down to integrated flux densities $\sim \, 2 \, \mu$Jy and $\sim \, 2.5 \, \mu$Jy respectively. 

The source counts are computed using the integrated flux densities (which means peak brightness for unresolved and integrated flux density for resolved), but sources are detected based on their measured peak brightness over the local noise level. The completeness of the source counts will thus depend both on the variation of the noise in the image and on the relation between integrated flux densities and peak brightness. In the following, we discuss these effects and how we correct for them in deriving the source counts.

\subsubsection{Multi-component sources}
\label{sec:Multi-component sources}
\begin{figure}
\centering
\includegraphics[width=\columnwidth]{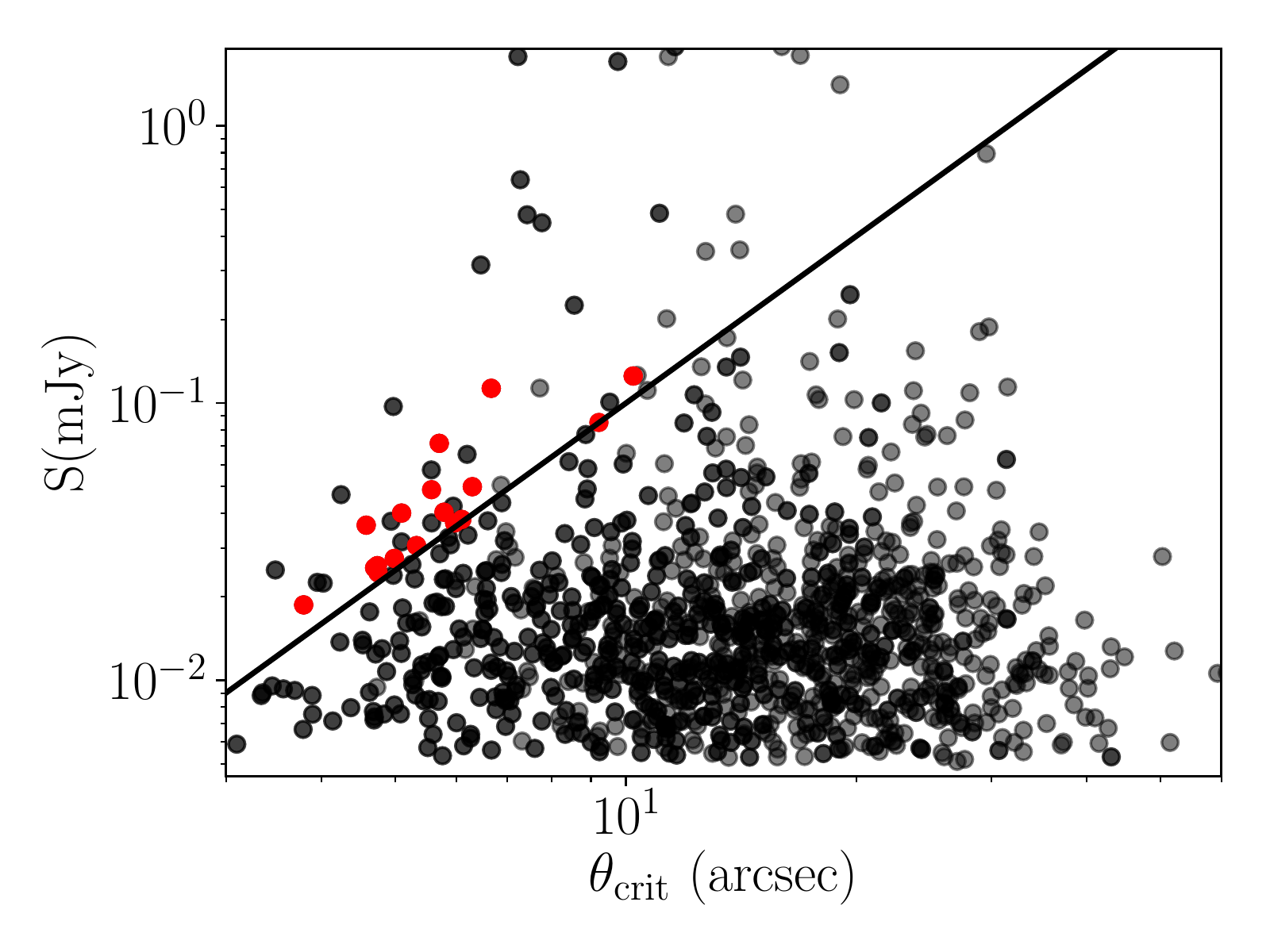}
\caption{Sum of the flux densities of nearest neighbor pairs as a function of their separation. Source pairs that have separation less then the critical value given by the solid line and flux densities that differ by less than a factor of 4 are considered as double source candidates. 17 sources are identified as multi-component sources and are indicated with red points.}
\label{fig:multicomponent_sources}
\end{figure}

In order to derive the source counts, we need to take into account that a single source may be made up of multiple components. For example, radio sources associated with radio galaxies can be made up of a nucleus with hot spots along, or at the end of, one or two jets. When jets are detected, it is relatively easy to recognize the components belonging to the same source. When a jet is missing, the radio-lobes are detected as two separated sources. We apply the statistical technique described by \cite{Magliocchetti_1998} and \cite{White_2012} to find these double component sources. We consider the separation of the nearest neighbour of each component and the summed flux density of the source and its neighbour. Multi-components are combined as single sources if the ratio of their flux densities is between 0.25 and 4 and their separation is less than a critical value dependent on their integrated flux density given by: 

\begin{equation}
\theta_{\rm crit} = 100\left(\frac{S_{\rm sum}}{10}\right)^{0.5}\, ,
\end{equation}

\noindent where $S_\text{sum}$ is in mJy and $\mathbf{\theta_{\rm crit}}$ is in arcsec. This maximum separation is shown in Fig. \ref{fig:multicomponent_sources}. We analyzed both the 10 and $3 \, \rm{GHz}$ image but only found multi-component sources for the $3 \, \rm{GHz}$ image. The $3 \, \rm{GHz}$ sources that meet both requirements are shown in Fig. \ref{fig:multicomponent_sources} as red circles. From this analysis we could identify 17 multi-component sources. See also Fig. \ref{fig:sources} for an example of a multi-component source.

\subsubsection{Completeness and reliability}
We consider a correction for the completeness of the catalog as derived in Section \ref{sec:Completeness}. Using the derived completeness corrections we can calculate the source counts ($n(S)$) in each flux density bin using

\begin{equation}
n(S) = \frac{1}{A}\sum^{N}_{{\rm i}=1}\frac{1}{C_{\rm compl}(S_{\rm i})} \, ,
\label{eq:n_1}
\end{equation}

\noindent where $A \, \sim \, 350 \, \text{arcmin}^2$ is the solid angle considered for the source counts, $N$ is the number of sources in the flux density bin and $C_{\rm compl}(S_{\rm i})$ is the derived completeness for the given flux density. 

Additionally, we make a correction for the reliability for the source counts at $3 \, \rm{GHz}$ by applying the false detection rate derived in Section \ref{sec:Reliability}. This correction acts in the opposite direction to the completeness correction and can be added to Equation \ref{eq:n_1}:

\begin{equation}
n(S) = \frac{1}{A}\sum^{N}_{{\rm i}=1}(1-F_{\rm false-det}(S_{\rm i})) \frac{1}{C_{\rm compl}(S_{\rm i})}\, ,
\end{equation}

\noindent where $F_{\rm false-det}(S_{\rm i})$ is the correction for the false detection rate for the given flux density bin. 

\subsubsection{Resolution bias}
\label{sec:resolution bias}
\begin{figure}
\centering
\includegraphics[width=\columnwidth]{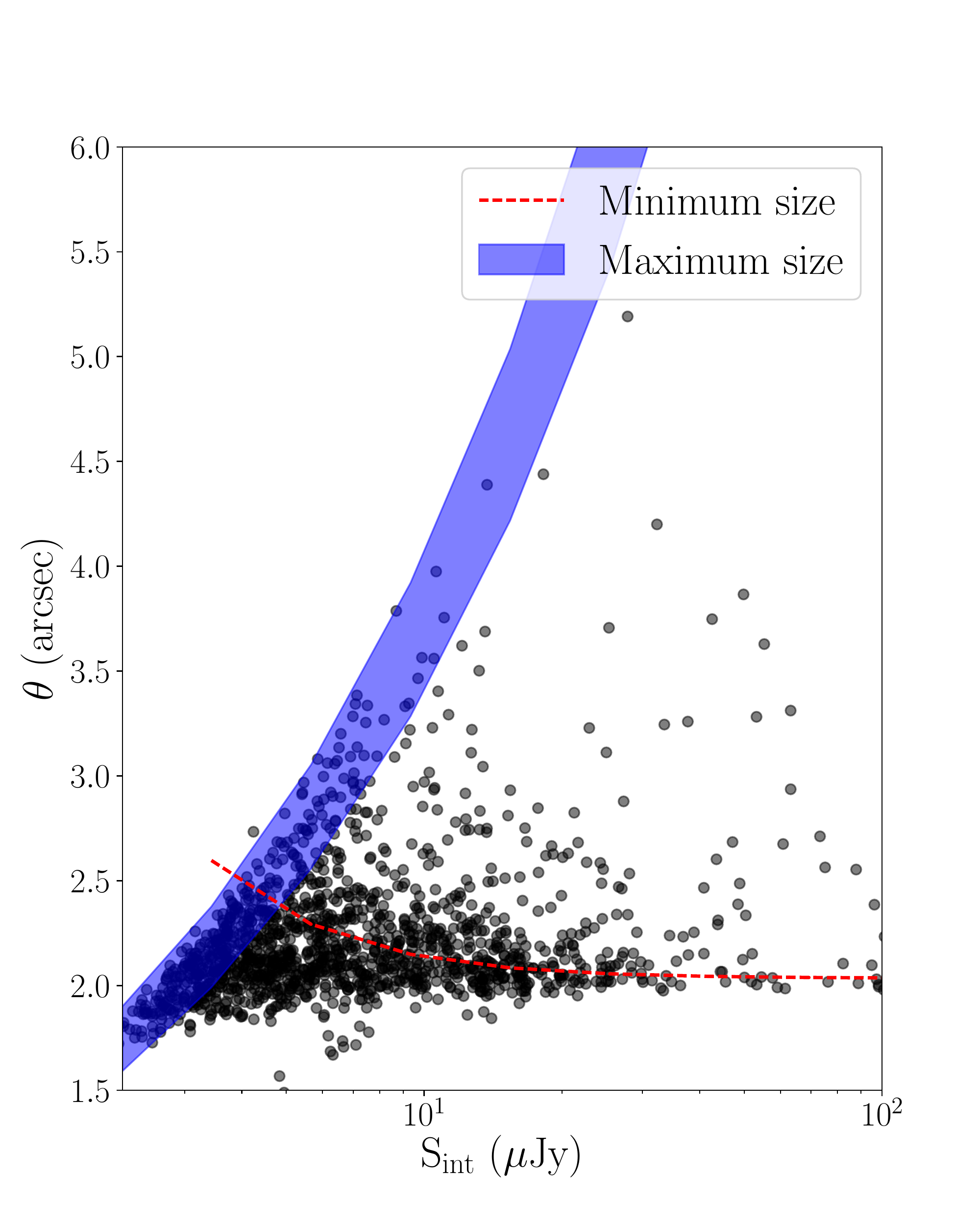}
\caption{The fitted angular size, $\theta$ (geometric mean), as a function of integrated flux density at $3 \, \rm{GHz}$. Unresolved sources fall below the red dotted line giving the minimum size of a source. The minimum size was derived using the envelope from Section \ref{sec:Resolved_sources}. The blue shaded region shows the maximum size a source of a given integrated flux density can have before dropping below the peak brightness detection threshold where the range reflects the range of r.m.s. noise in the $3 \, \rm{GHz}$ image.}
\label{fig:max_size}
\end{figure}

Sources in our image are found by identifying peaks of emission above a given threshold. This means that a resolved source of a given integrated flux density will be missed more easily than a point source of the same integrated flux density. This incompleteness is called the resolution bias and causes the number of sources to be underestimated, particularly near the detection limit of the survey. We correct for the resolution bias by adopting the analytic method used in \cite{Prandoni_2001} and \cite{Williams_2016}.

The following relation can be used to calculate the maximum angular size that a source can have and still be detected for a given integrated flux density:

\begin{equation}
S_{\rm int}/5\sigma = \frac{\theta_{\rm min}\theta_{\rm maj}}{b_{\rm min}b_{\rm maj}} \, ,
\label{eq:max_size}
\end{equation}

\noindent where $\theta_{\rm min}$ and $\theta_{\rm maj}$ are the fitted source axes as measured by PyBDSF, $b_{\rm min}$ and $b_{\rm min}$ are the synthesized beam axes and $5 \sigma$ is the peak brightness detection limit (where $\sigma$ is the local r.m.s. in the image). We use this relation to calculate the maximum size that a source can have and still be detected. This relation is shown in Fig. \ref{fig:max_size}, where the range reflects the range of r.m.s. noise in the image. Fig. \ref{fig:max_size} also shows the distribution of $\theta$, the geometric mean of the source major and minor axes, as a function of integrated flux density for all sources in the $3 \, \rm{GHz}$ catalog. Also shown is the minimum size a source can have before it is deemed to be unresolved. We use the maximum size to calculate the fraction of sources expected to be larger than this value, following Windhorst, Mathis \& Neuschaefer (1990), using

\begin{equation}
h(>\theta_{\rm max}) = \exp \left[- \ln (2) \left(\frac{\theta_\text{max}}{\theta_\text{med}}\right)^{0.62}\right] \, ,
\end{equation}

\noindent where $\theta_{\rm max}$ is the maximum angular size and $\theta_{\rm med}$ the median angular size. $\theta_{\rm max}$ can be calculated by rewriting Equation \ref{eq:max_size}: 

\begin{equation}
\theta_{\rm max} = \sqrt{b_{\rm min}b_{\rm maj} \times \frac{S_{\rm int}}{5\sigma}}
\end{equation}

\noindent We use two different versions of $\theta_{\rm med}$ for comparison; the first, given by \citealt{Windhorst_1990},

\begin{equation}
\theta_{\rm med} = 2(S_{\rm 1.4 GHz})^{0.3} \, \rm ^{\prime \prime} 
\end{equation}

\noindent with $S_{\rm 1.4 GHz}$ in mJy (flux densities are scaled from $10 \, \rm{GHz}$ and $3 \, \rm{GHz}$ to $1.4 \, \rm{GHz}$ using a spectral index of $-0.7$), and the second adopting a constant size of $0.35^{\prime \prime}$ below $1 \, \rm{mJy}$ (based on recent results from \citealt{Cotton_2018} and \citealt{Bondi_2018}) as follows:

\begin{equation}
  \theta_{\rm med}=\begin{cases}
    0.35 \, \rm ^{\prime \prime}, & \text{for  $S_{\rm 1.4 GHz} \, < \, 1$ mJy,}\\
    2(S_{\rm 1.4 GHz})^{0.3} \, \rm ^{\prime \prime}, & \text{otherwise}.
  \end{cases}
\end{equation}

We can calculate the resolution-bias correction factor to be applied to the source counts using:

\begin{equation}
c = 1/[1-h(>\theta_{\rm max})] \, .
\end{equation}

\begin{figure}
\centering
\includegraphics[width=\columnwidth]{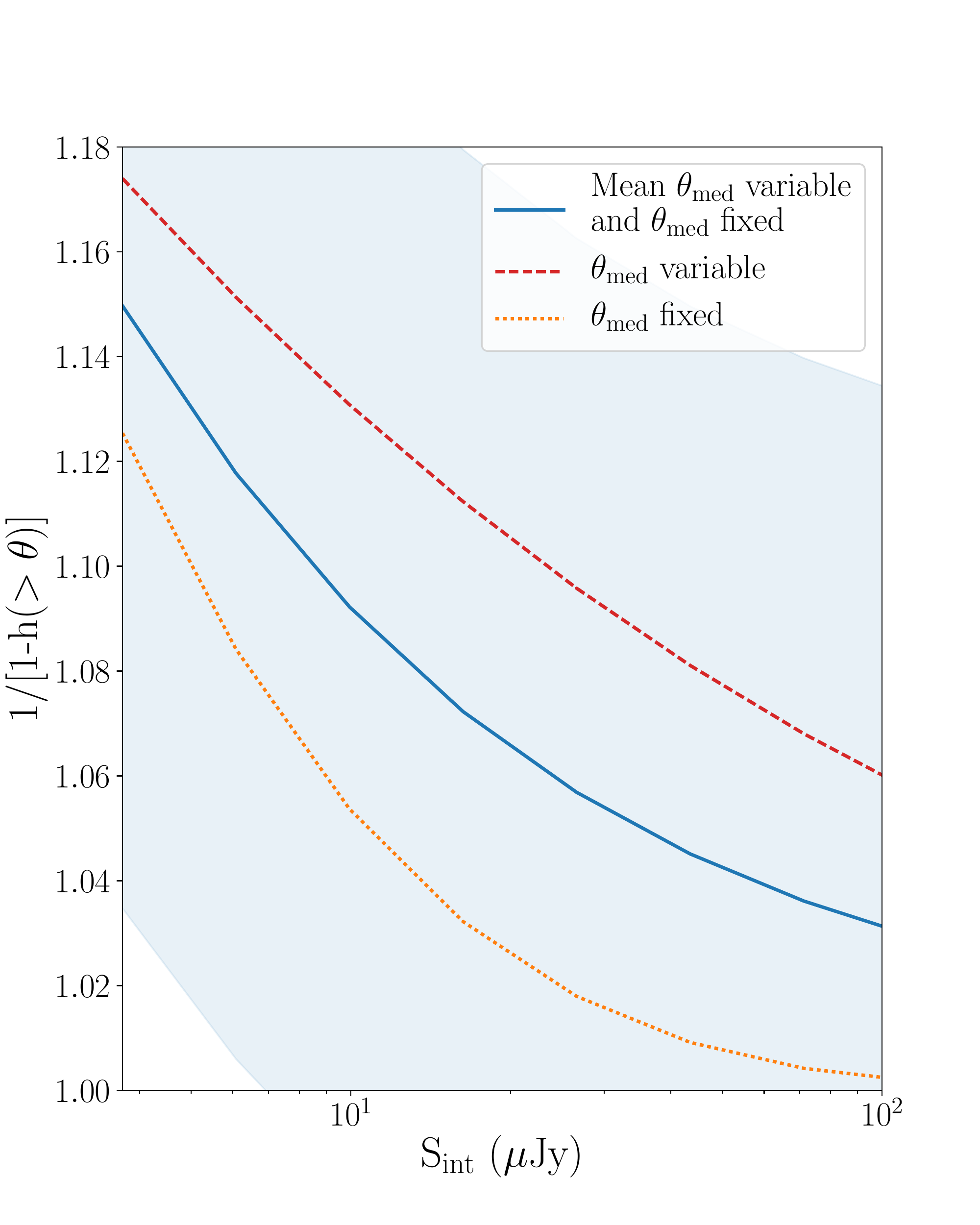}
\caption{Resolution bias correction $1/[1-h(>\theta_{\rm max})]$ for the fraction of sources with angular size larger than $\theta_{\rm max}$ at a given integrated flux density at $3 \, \rm{GHz}$. For the faintest sources, three curves are shown: the red dashed curve shows $\theta_{\rm med}$ as a function of $S_{\rm int}$, the orange dotted curve shows the result assuming $\theta_{\rm med} \, = \, 0.35^{\prime \prime}$ at these flux densities, and the blue solid curve shows the mean of the two curves. The range reflects the assumed 10 per cent uncertainty following \protect\cite{Windhorst_1990}. The resolution bias correction is found to be 1.15 for the lowest flux bin.}
\label{fig:bias_corr}
\end{figure}

The correction factors calculated using the two different size distributions are plotted as a function of the integrated flux density at $3 \, \rm{GHz}$ $S_{\rm int}$ in Fig. \ref{fig:bias_corr}. We use the mean of the two correction factors to correct the source counts in this paper, and use an identical procedure to obtain the correction at 10\,GHz. We use the uncertainty in the forms of $\theta_{\rm med}$ and in $\theta_{\rm max}$ to estimate the uncertainty in the resolution bias correction. We further include an overall 10 per cent uncertainty following \cite{Windhorst_1990} which dominates the error budget. We then determine the source counts, including a correction factor $c(S_i)$ for the resolution bias in each bin, via
\begin{equation}
n(S) = \frac{1}{A}\sum^{N}_{{\rm i}=1}(1-F_{\rm false-det}(S_{\rm i}))c(S_{\rm i}) \frac{1}{C_{\rm compl}(S_{\rm i})} \, .
\end{equation}

\subsubsection{The sub-$\mu$Jy source counts at $3 \, \rm{GHz}$ compared to observations}
\label{sec:source_counts_2}
\begin{figure*}
    \centering
    \subfigure[]
    {
        \includegraphics[width=1.0\columnwidth]{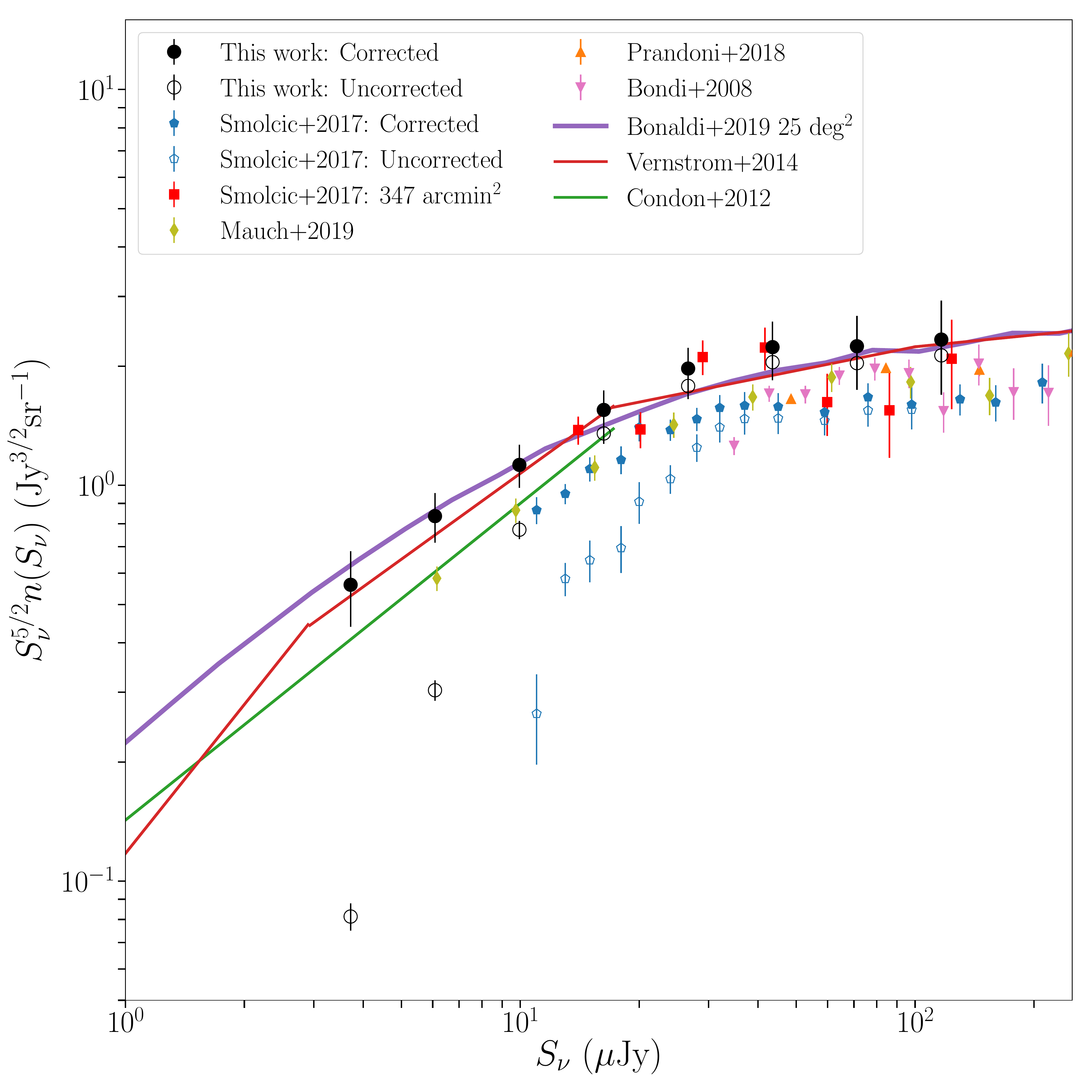}
        \label{fig:source_counts_cv}
    }
    \subfigure[]
    {
        \includegraphics[width=1.0\columnwidth]{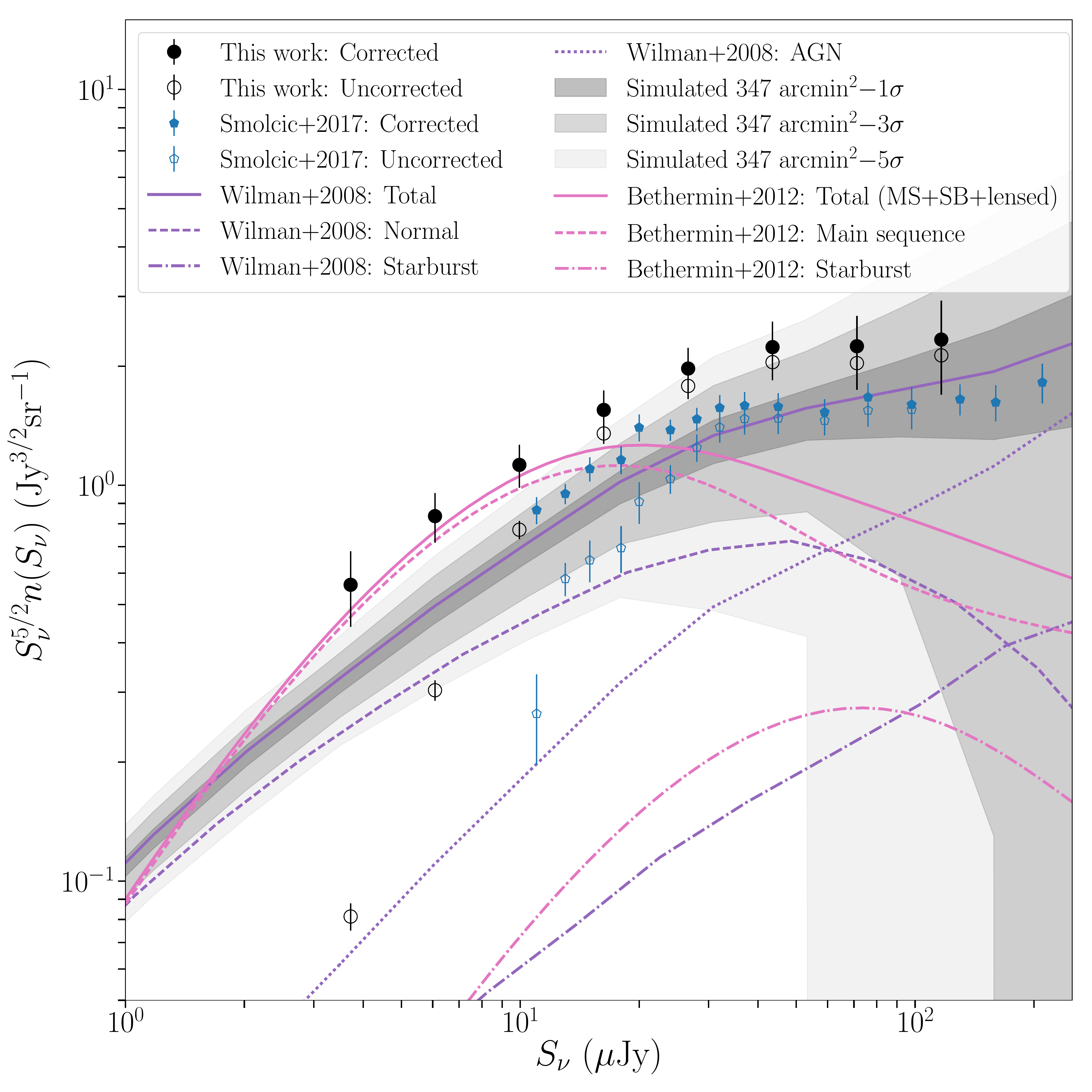}
        \label{fig:source_counts}
    }
    \caption{Euclidean-normalized radio source counts at $3 \, \rm{GHz}$ (filled black circles). Open black circles show the source counts without corrections. Error bars correspond to the Poisson error for the uncorrected points. For the corrected points the errors are propagated from the errors on the correction factors and the Poisson errors on the raw counts per bin. Corrected (closed) and uncorrected (open) VLA-COSMOS $3 \, \rm{GHz}$ measurements from \protect\cite{Smolcic_2017} are shown with blue points.
    Panel \subref{fig:source_counts_cv} also shows the number counts from $P(D)$ analysis by \protect\cite{Condon_2012} (green line) and \protect\cite{Vernstrom_2014} (red line). The purple line shows the results from the simulation of \protect\cite{Bonaldi_2019}. The red filled squares show the source counts for sources in the \cite{Smolcic_2017} catalog within the area of our pointing.
    Panel \subref{fig:source_counts_cv} also shows the results from $1.4 \, \rm{GHz}$ observations from \protect\cite{Bondi_2008} (filled pink downward triangles) and  \protect\cite{Prandoni_2018} (filled orange upward triangles) and $1.28 \, \rm{GHz}$ observations from \protect\cite{Mauch_2019} (filled olive diamonds). The flux densities are shifted to the $3 \, \rm{GHz}$ observed frame using a spectral index of $-0.84$. 
    Panel \subref{fig:source_counts} shows the comparison to the number counts from the simulation of \protect\cite{Wilman_2008} (purple lines) and \protect\cite{Bethermin_2012} (pink lines). Different line styles correspond to different source types as defined in the legend. The solid lines show the total source counts. The shaded region demonstrates the effect of cosmic variance and correspond to 1, 3 and 5 standard deviations in the source counts measurements of the simulations by \protect\cite{Wilman_2008}.}
\end{figure*}

The Euclidean normalized counts are shown in Fig. \ref{fig:source_counts_cv} and tabulated in Table \ref{tab:S_band_res}. Uncertainties on the final normalized source counts are propagated from the errors on the correction factors and the Poisson errors on the raw counts per bin. For the uncorrected data points, these errors are only the Poisson errors. Fig. \ref{fig:source_counts_cv} illustrates our results compared to other observational results, including the $3 \, \rm{GHz}$ surveys by \cite{Condon_2012}, \cite{Vernstrom_2014}, \cite{Smolcic_2017}. 

\begin{table}
\centering
\setlength{\tabcolsep}{0.39\tabcolsep}
\caption{Euclidean-normalized differential source counts for the $3 \, \rm{GHz}$ catalog.}
\label{tab:S_band_res}
\begin{threeparttable}
\begin{tabular}{*{8}{c}}
	\hline
	$\Delta S_{\nu}$ & $S_{\nu}$ & Counts & Error & N & $F_{\rm false-det}$ & $C_{\rm compl}$ & $c$ \\
    {[}$\mu$Jy{]} & {[}$\mu$Jy{]} & {[}Jy$^{1.5}$ & {[}Jy$^{1.5}$ &  &  &  &  \\
     & & sr$^{-1}${]} & sr$^{-1}${]} &  &  &  &  \\    
	\hline
2.82 -- 4.61 & 3.72 & 0.56 & 0.12 & 161 & 0.0 & 0.17 & 1.14 \\
4.61 -- 7.55 & 6.08 & 0.84 & 0.12 & 287 & 0.02 & 0.4 & 1.11 \\
7.55 -- 12.35 & 9.95 & 1.13 & 0.14 & 349 & 0.01 & 0.74 & 1.09 \\
12.35 -- 20.2 & 16.27 & 1.55 & 0.19 & 292 & 0.02 & 0.91 & 1.07 \\
20.2 -- 33.04 & 26.62 & 1.97 & 0.25 & 184 & 0.02 & 0.93 & 1.05 \\
33.04 -- 54.04 & 43.54 & 2.23 & 0.36 & 101 & 0.02 & 0.94 & 1.04 \\
54.04 -- 88.41 & 71.22 & 2.25 & 0.44 & 48 & 0.0 & 0.94 & 1.03 \\
88.41 -- 144.62 & 116.51 & 2.33 & 0.59 & 24 & 0.0 & 0.94 & 1.03 \\
        \hline
\end{tabular}
\begin{tablenotes}
\item \textbf{Notes.} The format is the following: Column (1): flux interval. Column (2): bin centre. Column (3): differential counts normalized to a non evolving Euclidean mode. Column(4): error on differential counts. Column (5): number of sources detected. Column (6): false detection rate. Column (7): completeness correction. Column (8): resolution bias correction. The listed differential counts were corrected for completeness and resolution bias ($C_{\rm compl}$ and $c$), as well as false detection fractions ($F_{\rm false-det}$), by multiplying the raw counts by the correction factor which is equal to $(c(S_{\rm i})*(1-F_{\rm false-det}(S_{\rm i})))/C_{\rm compl}(S_{\rm i})$. The source count errors take into account the Poissonian errors and completeness and bias correction uncertainties (see text for details).
\end{tablenotes}
\end{threeparttable}
\end{table}

\cite{Condon_2012} used the $P(D)$ (probability distribution of peak flux densities) analysis technique to statistically estimate the radio number counts down to $\sim \, 1 \, \mu\rm{Jy}$ from VLA $3 \, \rm{GHz}$ observations. $P(D)$ analyses tend to be less prone to resolution effects, as these study the statistical properties of sources well below the confusion limit of the survey. This approach results in statistical estimates of the source counts that are much fainter than the faintest sources that can be counted individually. 
\cite{Vernstrom_2014} also used the $P(D)$ analysis technique to re-analyse the $3 \, \rm{GHz}$ observations from \cite{Condon_2012} and estimate the radio number counts down to $\sim \, 50 \, \rm{nJy}$. They used a different approach from \cite{Condon_2012} that allowed for more flexibility in accurately modelling the true source counts. Our completeness corrected points are slightly higher than the \cite{Condon_2012} results and are more consistent with the \cite{Vernstrom_2014} results. 

\cite{Smolcic_2017} derived source counts from $3 \, \rm{GHz}$ observations of the two square degree COSMOS field from the VLA-COSMOS $3 \, \rm{GHz}$ Large Project. For comparison we show both the uncorrected number counts and those corrected for completeness, resolution bias and false detections. Our completeness corrected points consistently lie a factor $\sim \, 1.4$ above the source counts from \cite{Smolcic_2017}.

One possible cause for the offset between our source counts and those from \cite{Smolcic_2017} is an underestimated correction for the resolution bias. In particular, our survey has a lower resolution than the \cite{Smolcic_2017} observations (2$^{\prime \prime}$ versus 0.75$^{\prime \prime}$) and is therefore less likely to miss sources due to the resolution bias (see also the small correction derived in Section \ref{sec:resolution bias}). 
\cite{Smolcic_2017} performed extensive Monte Carlo simulations to account for the resolution bias. They modeled the intrinsic angular sizes of mock sources using a simple power-law parametrization of the angular size distribution as a function of their total flux density, as derived by \cite{Bondi_2008}. They further applied a minimum angular size for faint mock sources to ensure the modelled angular size distribution matched the observed size distribution. We investigated whether the difference in resolution bias correction methods between \cite{Smolcic_2017} and the one used in this work (described in Section~\ref{sec:resolution bias}) is significant, finding that when applied to the same data, our method results in a higher normalization by a factor of $\sim$\,1.1 (i.e., $\sim$25\% of the observed difference). Therefore, the difference in resolution bias correction method may partly explain the offset between our number counts and those reported in \cite{Smolcic_2017}.

To attempt to gain some additional insight, we also compare our results with those from recent $1.4 \, \rm{GHz}$ (or similar) observations in Fig. \ref{fig:source_counts_cv}.
We scale those flux densities to the $3 \, \rm{GHz}$ observed frame assuming a spectral index of $-0.84$.\footnote{\cite{Smolcic_2017} found this spectral index by performing a Gaussian fit to the spectral index distribution of 3 GHz sources which were also detected at 1.4 GHz. This spectral index is different from the value derived using a survival analysis which takes non-detections into account. The difference can be explained by selection effects as discussed in \cite{Condon_1984_b}.} \cite{Bondi_2008} derive their counts in the inner 1 deg$^2$ region of the COSMOS field from the VLA-COSMOS $1.4 \, \rm{GHz}$ Large Project (\citealt{Schinnerer_2007}) catalog, which required a somewhat uncertain correction for the effect of bandwidth smearing, while \cite{Prandoni_2018} derived their counts from $1.4 \, \rm{GHz}$ mosaic observations obtained with the Westerbork Synthesis Radio Telescope (WSRT). Because the latter cover an area of 6.6 square degrees, the source catalogue contains $\sim \, 6000$ sources (note that the errorbars in Fig. \ref{fig:source_counts_cv} are smaller than the symbols). Finally, \cite{Mauch_2019} used the $P(D)$ analysis technique to derive source counts from $0.25 \, \mu\rm{Jy}$ to $10 \, \mu\rm{Jy}$ using confusion-limited  $1.28 \, \rm{GHz}$ GHz MeerKAT observations. They further derive the source counts between $10 \, \mu\rm{Jy}$ and $2.5 \, \rm{mJy}$ from individual detected sources. The direct source counts (scaled to $3 \, \rm{GHz}$) are shown in Fig. \ref{fig:source_counts_cv}. 

With the assumed spectral index of $\alpha=-0.84$, we find that the source counts of \cite{Bondi_2008}, \cite{Prandoni_2018}, and \cite{Mauch_2019} generally fall between the \cite{Smolcic_2017} counts and those derived here, and are on average lower than our source counts (particularly the \cite{Mauch_2019} counts, which also extend to the lowest flux densities). As already noted by \cite{Smolcic_2017}, 
the comparison between the 1.4 and 3 GHz source counts is complicated by the potentially overly simplistic scaling of the 3 GHz counts to 1.4 GHz using a single spectral index value (in addition to the varying resolution bias and bandwidth smearing effects present). We therefore investigate one final effect that may influence the measured source counts in the following section.

\subsubsection{Cosmic variance}
\label{sec:Cosmic_variance}
A final source of uncertainty that may affect the measured source counts is
cosmic variance. In particular, if we observe overdensities in our single pointing, the resulting number counts will be higher than the number counts averaged over a larger area (as in \citealt{Smolcic_2017}).

\cite{Heywood_2013} developed a method to assess the influence of source clustering on radio source counts, by extracting a series of independent samples from the models of \cite{Wilman_2008}. We used a similar approach with both the \cite{Wilman_2008} simulation and the \cite{Bonaldi_2019} simulation, which are further discussed in the next section, to estimate the uncertainty induced by sample variance on a survey with properties similar to ours. Following \cite{Heywood_2013}, we extract multiple  non-overlapping sky patches with areas of $0.31 \, \times \, 0.31 \, \rm{deg}^2$ from the \cite{Wilman_2008} simulation (comparable to the effective area of our observations). We used a simulated area of $4 \, \times \, 4$ deg$^2$. This process resulted in 169 source catalogs. For each of these simulated source subsets, we compute the Euclidean normalized differential source counts. Fig. \ref{fig:source_counts} shows the mean value of the simulated counts from the independent distributions in each bin with the solid purple line. 
The shaded regions surrounding this correspond to one, three and five times the standard deviation of the count measurements. Fig. \ref{fig:source_counts} shows that count fluctuations found in our observed survey area are significant enough to dominate the observed scatter at flux densities above $100 \, \mu$Jy, and contribute significantly below this.

A similar approach was used to produce the shaded regions in Fig. \ref{fig:source_counts_bonaldi}. Here, we extract multiple non-overlapping sky patches with areas of $0.31 \, \times \, 0.31 \, \rm{deg}^2$ from the simulation of \cite{Bonaldi_2019}. We used a simulated area of 25 deg$^2$, resulting in 289 source catalogs. For each of these simulated source subsets, we compute the Euclidean normalized differential source counts. The shaded regions in Fig. \ref{fig:source_counts} and Fig. \ref{fig:source_counts_bonaldi} show that the offset between our number counts and the \cite{Smolcic_2017} counts could be at least partly explained by cosmic variance. 

We are able to further test the magnitude of cosmic variance by using the fact that the \cite{Smolcic_2017} counts are derived from an area that also covers our pointing. By counting the sources in the \cite{Smolcic_2017} catalog within the area of our pointing and assuming the completeness corrections described in \cite{Smolcic_2017}, we can compare their source density directly with our number counts. This comparison is shown in Fig. \ref{fig:source_counts_cv}. The number counts in the \cite{Smolcic_2017} catalog over our pointing are noisy due to the small number statistics in this (shallower) sample, but indicate a slightly higher source density (on average) within the area of our survey. We therefore conclude that the systematic offset observed between our 3 GHz number counts and the 3 GHz direct detections from \cite{Smolcic_2017} can most likely be explained by a combination of cosmic variance in our pointing and an underestimated resolution bias correction in \cite{Smolcic_2017}.

In summary, we find that our 3 GHz source counts agree with those at both 3 and 1.4 GHz within the various uncertainties, while extending down to typically lower flux densities than previous direct detections. In the following section, we will therefore explore the implications of our derived source counts on the modeling of the ultra-faint radio population.


\subsubsection{The sub-$\mu$Jy source counts at $3 \, \rm{GHz}$ compared to simulations}
\label{sec:source_counts}

In Fig. \ref{fig:source_counts} we compare our number counts with semi-analytical models from \cite{Wilman_2008} and \cite{Bethermin_2012}. \cite{Wilman_2008} developed a semi-empirical simulation that used observed and extrapolated luminosity functions to populate an evolving dark matter skeleton with various galaxy types (normal, starburst and AGN). The adopted distribution of sources on the underlying dark matter density was chosen to reflect their measured large-scale clustering. The \cite{Bethermin_2012} model uses two main ingredients to predict the number counts: the evolution of main-sequence galaxies and starburst galaxies based on the 2-Star-Formation-Mode framework of \cite{Sargent_2012} and main-sequence and starburst SED's to predict the shape of the IR luminosity function at $z \, \leq \, 2$. \cite{Bethermin_2012} also included dust attenuation and strong lensing in their model. To get to $1.4 \, \rm{GHz}$ radio source counts, they assumed a non-evolving IR-radio correlation and a synchrotron spectral slope of $\alpha \, = \, -0.8$.

Our derived source counts deviate from those predicted by the \cite{Wilman_2008} model. In particular, they tend to be higher than the predicted model from $\sim \, 40 \, \mu$Jy downward (and we note that the \cite{Smolcic_2017} counts are also systematically above this model at low flux densities despite a possibly underestimated resolution bias correction -- see Section~\ref{sec:source_counts_2}).
On the other hand, our 3 GHz source counts are in good agreement with the \cite{Bethermin_2012} model below $\sim10 \, \mu$Jy. The AGN population is not included in the \cite{Bethermin_2012} model and this explains the decrease in the source counts above $20 \, \mu$Jy as AGN start contributing significantly at these flux densities. Below $20 \, \mu\rm{Jy}$ the star-forming population dominates the number counts, as can be seen from Fig. \ref{fig:source_counts_bonaldi} and Fig. \ref{fig:source_counts}. \cite{Bethermin_2012} model this population using a framework that uses more recent results to predict the LF of SFGs at $z \, \leq \, 2$ compared to the \cite{Wilman_2008} model. \cite{Bethermin_2012} use a combination of mid-IR and UV data (\citealt{Daddi_2007, Elbaz_2007, Noeske_2007}), far-IR data (\citealt{Elbaz_2011, Pannella_2015}), and radio continuum imaging (\citealt{Karim_2011}). \cite{Wilman_2008} assume pure luminosity evolution out to $z \, = \, 1.5$ of the local LF derived from the IRAS 2-Jy sample by \cite{Yun_2001}. Our results suggest that the \cite{Wilman_2008} model could be improved by using the most recent observations to derive a model for the LF for SFGs. 

\cite{Bonaldi_2019} developed an improved simulation in the spirit of that of \cite{Wilman_2008}: the Tiered Radio Extragalactic Continuum Simulation (T-RECS). \cite{Bonaldi_2019} model the radio sky in terms of two main populations, AGNs and SFGs, and use these to populate an evolving dark matter skeleton. To describe the cosmological evolution of the LF of AGNs, they adopted an updated model of \cite{Massardi_2010}, with the revision of \cite{Bonato_2017}. The LF of SFGs is derived from the evolving SFR function as the radio continuum emission is correlated with the SFR. The SFR function gives the number density of galaxies per logarithmic bin of SFR at a given redshift $z$. The evolution of the synchrotron-SFR relation accounts for the evolving radio-FIR correlation. \cite{Bonaldi_2019} use the SFR function with redshift derived by \cite{Cai_2013} and \cite{Cai_2014} with an extension derived by \cite{Mancuso_2015}. The resulting SFR function extends to $z \, \sim \, 10$ and includes effects from strong gravitational lensing. 

We compare our $3 \, \rm{GHz}$ source counts to the simulation from \cite{Bonaldi_2019}, using their medium tier catalog of 25 deg$^{2}$ containing $3 \, \rm{GHz}$ flux densities. Our number counts, shown in Fig. \ref{fig:source_counts_bonaldi}, match the total number counts from the simulations. Fig. \ref{fig:source_counts_bonaldi} also shows the comparison between our source counts and those from the model by \cite{Mancuso_2017}. This model includes three populations: radio loud (RL) AGN, radio quiet (RQ) AGN and SFG. Using the same prescriptions as \cite{Bonaldi_2019}, they convert SFRs into 10 and $3 \, \rm{GHz}$ luminosities. The SFR function used is, however, slightly different from the function used in \cite{Bonaldi_2019} (see Section \ref{sec:source_counts_10GHz}). The RL AGNs are modeled the same way as in \cite{Bonaldi_2019}. To model RQ AGNs, the SFR function is mapped into the AGN luminosity function. The obtained AGN bolometric luminosity is subsequently converted to X-ray luminosity. 
The AGN radio power is then derived using the observed relation between rest-frame X-ray and 1.4 GHz radio luminosity for RQ AGNs.

The \cite{Mancuso_2017} model uses an intrinsic SFR function and therefore claims to be less sensitive to dust extinction effects. The derived intrinsic SFR function implies a heavily dust-obscured galaxy population at high redshifts ($z \, > \, 4$) and at large SFRs (10$^{2}$ M$_{\odot}$ yr$^{-1}$) (\citealt{Mancuso_2016}). Our number counts, shown in Fig. \ref{fig:source_counts_bonaldi}, fall well above the total number counts predicted by \cite{Mancuso_2017} at the faint end. 
In addition, there is a clear difference between the \cite{Bonaldi_2019} simulations and the \cite{Mancuso_2017} model, which is surprising as they use similar assumptions. The difference between the two simulations is further discussed in Section \ref{sec:source_counts_10GHz}.

To summarize, \cite{Wilman_2008} model SFGs assuming pure luminosity evolution out to $z \, = \, 1.5$ for the local LF, while thereafter no further evolution is assumed. This model reproduces the sources in the local universe as it matches the observed number density, with these sources dominating the flux densities at $> \, 2 \, \rm{mJy}$. However, for flux densities $< \, 40 \, \mu\rm{Jy}$ we find that the \cite{Bethermin_2012} model and the \cite{Bonaldi_2019} simulation are in better agreement with our observations. These models are able to produce the observed excess with respect to the source counts predicted by the \cite{Wilman_2008} simulation. Our number counts support the steeper LF evolution for SFGs that is used in these models.

\subsubsection{The sub-$\mu$Jy source counts at $10 \, \rm{GHz}$}
\label{sec:source_counts_10GHz}

\begin{figure*}
    \centering
    \subfigure[]
    {
        \includegraphics[width=1.0\columnwidth]{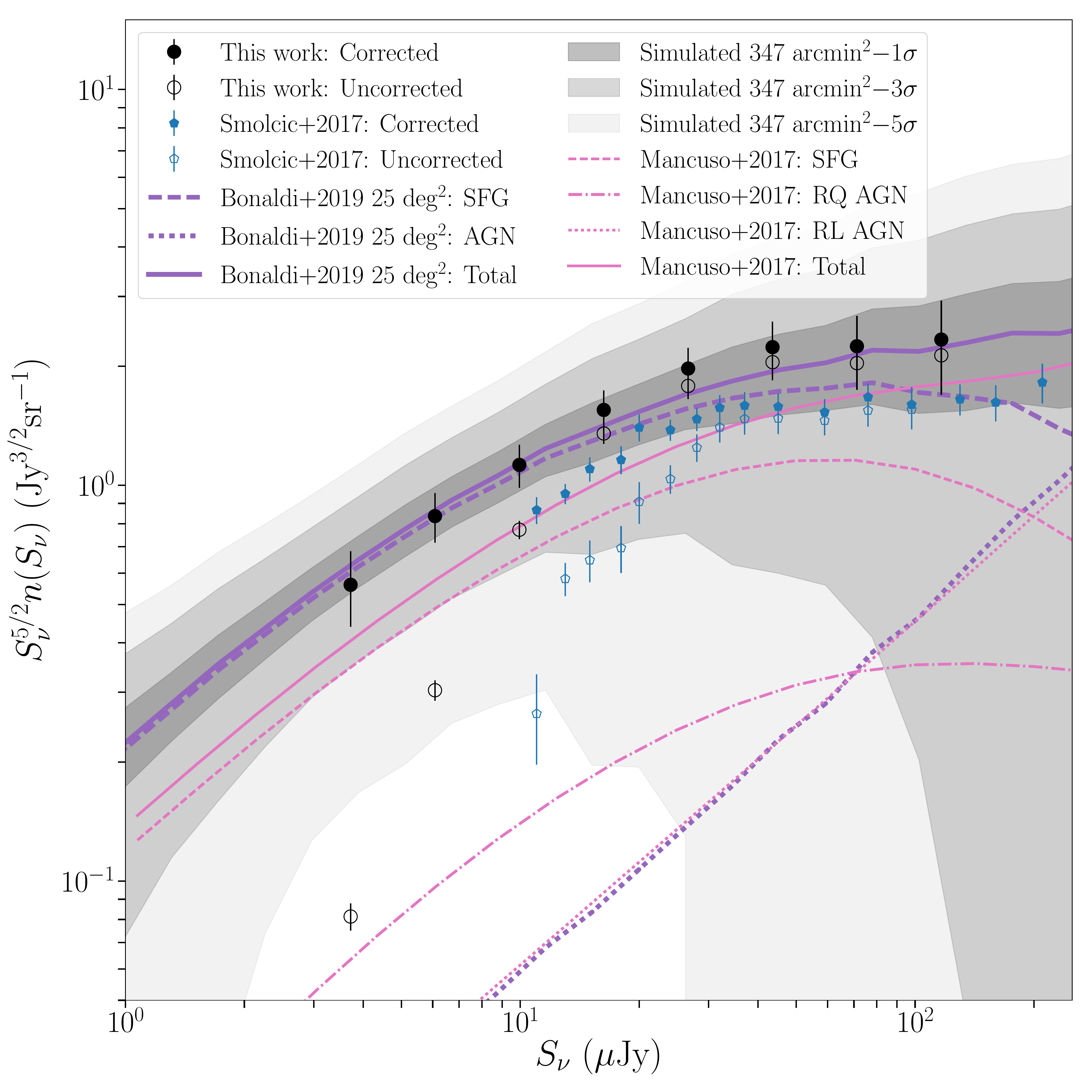}
        \label{fig:source_counts_bonaldi}
    } 
    \subfigure[]
    {
        \includegraphics[width=1.0\columnwidth]{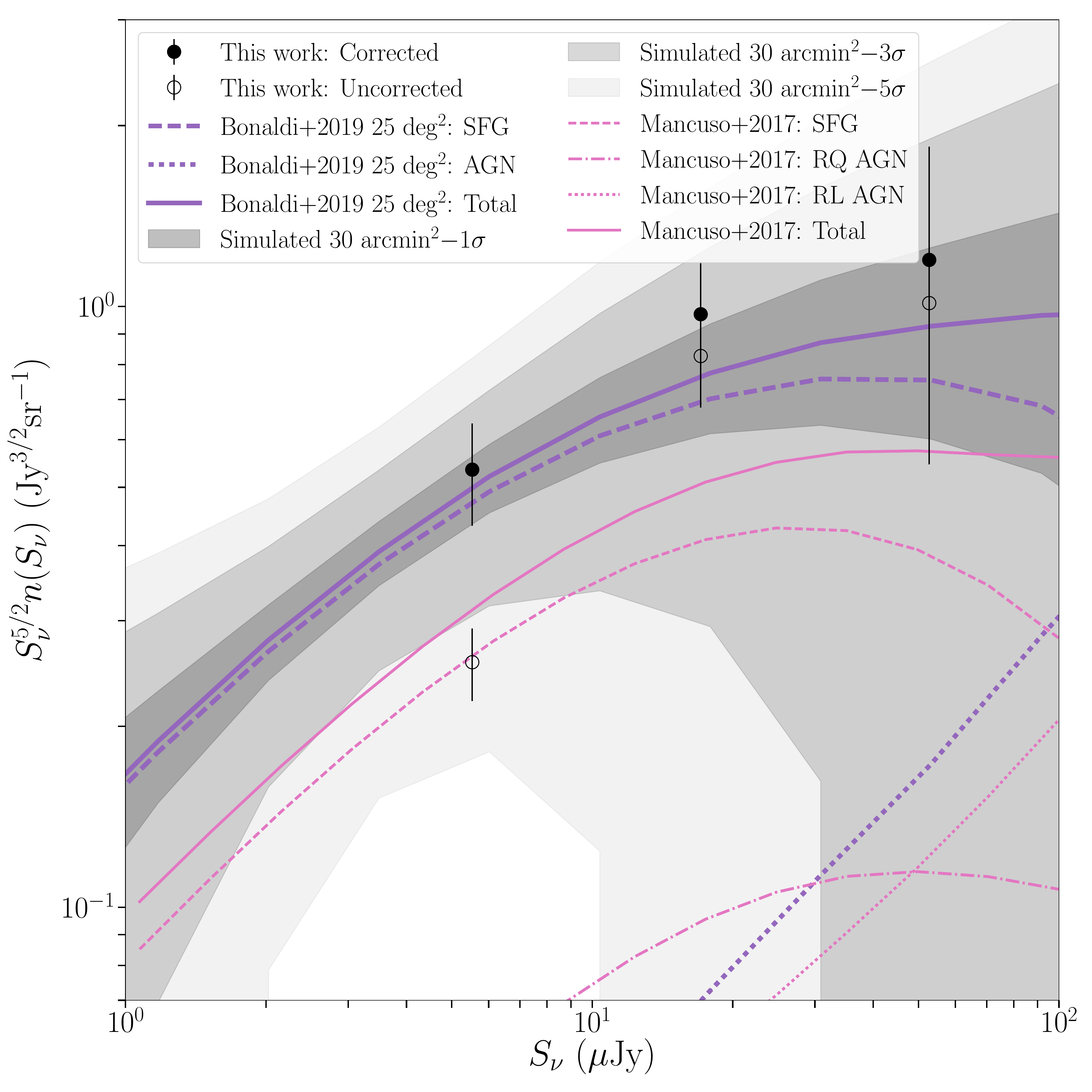}
        \label{fig:source_counts_bonaldi_10GHz}
    }
    \caption{Euclidean-normalized radio source counts at $3 \, \rm{GHz}$ (Panel \subref{fig:source_counts_bonaldi}) and $10 \, \rm{GHz}$ (Panel \subref{fig:source_counts_bonaldi_10GHz}). The filled black circles show the source counts from COSMOS-XS. Open black circles show the source counts without corrections. Error bars correspond to the Poisson error for the uncorrected points. For the corrected points the errors are propagated from the errors on the correction factors and the Poisson errors on the raw counts per bin. Corrected (closed) and uncorrected (open) VLA-COSMOS $3 \, \rm{GHz}$ measurements from \protect\cite{Smolcic_2017} are shown with blue points. The number counts from the simulation of \protect\cite{Bonaldi_2019} and predictions from \protect\cite{Mancuso_2017} are shown in purple and pink lines, respectively. Different line styles corresponding to different source types as defined in the legend. The solid lines show the total source counts from the simulation. The shaded regions demonstrate the effect of cosmic variance and correspond to 1, 3 and 5 standard deviations in the source counts measurements of the simulations by \protect\cite{Bonaldi_2019}.}
\end{figure*}

The Euclidean normalized counts at $10 \, \rm{GHz}$ are shown in Fig. \ref{fig:source_counts_bonaldi_10GHz}. Uncertainties on the final normalized source counts are propagated from the errors on the correction factors and the Poisson errors on the raw counts per bin. The source counts are tabulated in Table \ref{tab:X_band_res}.

\begin{table}
\centering
\setlength{\tabcolsep}{0.6\tabcolsep}
\caption{Euclidean-normalized differential source counts for the $10 \, \rm{GHz}$ catalog.}
\label{tab:X_band_res}
\begin{threeparttable}
\begin{tabular}{*{8}{c}} 
	\hline
	$\Delta S_{\nu}$ & $S_{\nu}$ & Counts & Error & N & $F_{\rm false-det}$ & $C_{\rm compl}$ & $c$ \\
    {[}$\mu$Jy{]} & {[}$\mu$Jy{]} & {[}Jy$^{1.5}$ & {[}Jy$^{1.5}$ &  &  &  &  \\
     & & sr$^{-1}${]} & sr$^{-1}${]} &  &  &  &  \\    
	\hline
2.71 -- 8.36 & 5.53 & 0.53 & 0.1 & 52 & 0.0 & 0.53 & 1.1 \\
8.36 -- 25.8 & 17.08 & 0.97 & 0.22 & 31 & 0.0 & 0.9 & 1.06 \\
25.8 -- 79.66 & 52.73 & 1.2 & 0.65 & 7 & 0.0 & 0.88 & 1.04 \\
    \hline
\end{tabular}
\begin{tablenotes}
\item \textbf{Notes.} The format is the following: Column (1): flux interval. Column (2): bin centre. Column (3): differential counts normalized to a non evolving Euclidean mode. Column(4): error on differential counts. Column (5): number of sources detected. Column (6): false detection rate. Column (7): completeness correction. Column (8): resolution bias correction. The listed differential counts were corrected for completeness and resolution bias ($C_{\rm compl}$ and $c$), as well as false detection fractions ($F_{\rm false-det}$), by multiplying the raw counts by the correction factor which is equal to $(c(S_{\rm i})*(1-F_{\rm false-det}(S_{\rm i})))/C_{\rm compl}(S_{\rm i})$. The source count errors take into account the Poissonian errors and completeness and bias correction uncertainties (see text for details).
\end{tablenotes}
\end{threeparttable}
\end{table}

With an observing frequency of $10 \, \rm{GHz}$, we measure flux densities closer to the rest-frame frequencies $\nu \, \geq \, 30 \, \rm{GHz}$ where the total radio emission is dominated by free-free radiation (e.g \citealt{Condon_1992, Murphy_2011, Klein_2018}). The calibration of the free-free radiation and SFR in the \cite{Bonaldi_2019} simulations follows \cite{Mancuso_2015} and \cite{Murphy_2011}. The \cite{Mancuso_2017} model used the same calibration for SFGs.

We can compare the $10 \, \rm{GHz}$ source counts to the simulation from \cite{Bonaldi_2019} by interpolating between the flux densities of their simulated sources given at 9.2 GHz and 12.5 GHz. Our number counts fall slightly above the total number counts from the simulations. The discrepancy is however within the $\sigma$ derived for the cosmic variance, indicated through the gray shaded regions in Fig. \ref{fig:source_counts_bonaldi_10GHz}. The shaded regions are derived in the same way as described in Section \ref{sec:Cosmic_variance} by extracting sky patches of $0.09 \, \times \, 0.09 \, \rm{deg}^2$ from the \cite{Bonaldi_2019} simulations. This process results in 3025 source catalogs. 

Our number counts are systematically higher than those predicted by \cite{Mancuso_2017} at $10 \, \rm{GHz}$, especially at the faint end. This would indicate that \cite{Mancuso_2017} underestimates the flux density produced at $10 \, \rm{GHz}$, especially by SFGs, as these contribute most at low flux densities as can be seen in Fig. \ref{fig:source_counts_bonaldi_10GHz}.

As seen in Fig. \ref{fig:source_counts_bonaldi} and Fig. \ref{fig:source_counts_bonaldi_10GHz}, the \cite{Bonaldi_2019} and \cite{Mancuso_2017} simulations predict roughly the same general shape of the number counts, but they predict a different normalization. The \cite{Mancuso_2017} model includes the specific modeling of RQ AGN. In the \cite{Bonaldi_2019} simulations, RQ AGN are not specifically modeled, and \cite{Bonaldi_2019} mention that RQ AGNs would contribute part of the flux density of those sources that, in T-RECS, are modelled as SFGs. This can therefore not explain the difference in normalization between the total source counts. 

Instead, the difference in normalization may be partially explained by how the two models treat the evolution of the radio-FIR correlation. \cite{Bonato_2017} already showed that the radio source counts support an evolving radio-FIR correlation. In their study, the source counts found without an evolving radio-FIR correlation are found to be substantially below the observational determinations. The \cite{Mancuso_2017} model does not include the increase of the ratio between sychrotron luminosity and far-IR luminosity as was recently reported by \cite{Delhaize_2017} and \cite{Magnelli_2015} and falls thus below the observed source counts. \cite{Bonaldi_2019} evolve the synchrotron-SFR relation to account for the evolving radio-FIR correlation, yielding a very good fit to the observational estimates of the radio luminosity function of SFGs (\citealt{Novak_2017}).

In addition to their different treatment of the radio-FIR correlation, both the \cite{Mancuso_2017} and \cite{Bonaldi_2019} studies also use a slightly different SFR function to make number counts predictions. 
In particular, both studies use a smooth, analytic representation of the SFR function derived from IR and dust-extinction-corrected UV, Ly$\alpha$ and H$\alpha$ data.
However, 
\cite{Mancuso_2017} use a standard Schechter shape characterized by three evolving parameters: the normalization, the characteristic SFR and the faint-end slope. Meanwhile, 
\cite{Bonaldi_2019} use the modified Schechter function introduced by \cite{Aversa_2015}, with two evolving characteristic slopes. Our observed number counts are in better agreement with the SFR function used by \cite{Bonaldi_2019}, which was derived by \cite{Mancuso_2015}, although the function used in \cite{Mancuso_2017} was derived using the most recent IR and UV data.

In summary, as seen in Sections \ref{sec:source_counts} and \ref{sec:source_counts_10GHz}, models 
typically 
assume either 
an evolving LF or an evolving SFR function to make number counts predictions at the $\mu$Jy level. We find that 
our source counts are in agreement with the \cite{Bonaldi_2019} model that uses an evolving radio-FIR relation and the modified Schechter function to describe the SFR function. We also find that the source counts predictions from the evolving LF for SFGs used by \cite{Bethermin_2012} result in a 
good match with the observed number counts below $10 \, \mu$Jy. 

The relation between the LF and the SFR functions will be investigated in a future publication. The luminosity-SFR relation is of crucial importance for measuring the cosmic star-formation history, which will also be studied in a future publication. In addition, we can utilize the multi-wavelength information available in the COSMOS region to determine the composition of the faint radio population responsible for the measured number counts. This will be fully discussed and presented in a companion paper (Algera et al. 2020). 

\section{Summary \& Conclusions}
\label{sec:Conclusions}
In this paper, we presented the details of the COSMOS-XS survey: an ultra--deep pointing in the COSMOS field at 10 and 3 GHz with the Karl G. Jansky VLA. 
The final 10 and $3 \, \rm{GHz}$ images have a resolution of $\sim \, 2^{\prime \prime}$ and reach a median r.m.s. of $0.41 \, \mu \rm{Jy} \, \rm{beam}^{-1}$ and $0.53 \, \mu\rm{Jy} \, \rm{beam}^{-1}$, respectively. The two radio catalogs contain sources detected within 20\% of the peak primary beam sensitivity with peak brightness exceeding 5$\sigma$. The $10 \, \rm{GHz}$ catalog contains 91 sources and the $3 \, \rm{GHz}$ catalog contains 1498 sources. 

Comparing the positions of our $3 \, \rm{GHz}$ sources with those from the high-resolution VLBA imaging at $1.4 \, \rm{GHz}$, we estimated that the astrometric uncertainties are negligible. Completeness corrections are calculated using Monte Carlo simulations whereby the effect of the primary beam is taken into account. We find that the completeness is above $\sim \, 85$\% at $10 \, \mu$Jy/$15 \, \mu$Jy for the $10 \, \rm{GHz}$/$3 \, \rm{GHz}$ source catalogs, respectively. 
We correct for the resolution bias by utilising an analytic method as used in \cite{Prandoni_2001} and \cite{Williams_2016}.

The deep 10 and $3 \, \rm{GHz}$ observations enable us to investigate the Euclidean normalized source counts down to the $\mu$Jy level. Our corrected radio counts at $3 \, \rm{GHz}$ with direct detections down to $\sim \, 2.8 \, \mu$Jy are consistent within the uncertainties with other results at 3 and 1.4 GHz, but extend to fainter flux densities than previous direct detections.

In comparison to simulations developed in the framework of the SKA Simulated Skies project (\citealt{Wilman_2008}),
our source counts show an excess which cannot be accounted for by cosmic variance. Our measured source counts suggest a steeper LF evolution for SFGs for low flux densities ($< \, 2 \, \rm{mJy}$). The T-RECS simulations (\citealt{Bonaldi_2019}) and the \cite{Bethermin_2012} model both produce this steeper evolution and are in better agreement with our measured source counts.

Our corrected radio counts at $10 \, \rm{GHz}$ with direct detections down to $\sim \, 2.7 \, \mu$Jy are systematically higher than those predicted by the T-RECS simulations (\citealt{Bonaldi_2019}), but we demonstrate that this falls within the expected variations from cosmic variance. The more significant offset observed between our counts and \cite{Mancuso_2017} support an  evolving  radio-FIR  relation  and  a  modified  Schechter  function  to  describe the SFR  function, as done in \cite{Bonaldi_2019}. 

The COSMOS-XS survey is one of the deepest radio continuum surveys to-date, providing valuable information for next generation facilities, such as the ngVLA and SKA, which will achieve much deeper imaging. These radio data, in conjunction with the vast panchromatic COSMOS data sets, will allow us to study the dust-unbiased star-formation history, place critical constraints on dust attenuation and conduct a study of the long-wavelength spectra of faint radio sources.

\section*{Acknowledgements}
We thank the anonymous referee for helpful comments and suggestions that significantly improved this work. We thank Preshanth Jagannathan, Huib Intema and Reinout van Weeren for the useful discussions that helped improve the imaging with \texttt{WSclean}, Chris Carilli for the useful discussions while preparing the proposal and for his help planning the observations, Mara Salvato for providing us with the COSMOS spectroscopic master catalog, Ian Smail for his comments, Andrea Lapi for providing the evolutionary track shown in Fig. \ref{fig:source_counts_bonaldi} and Fig. \ref{fig:source_counts_bonaldi_10GHz}, which refers to the model discussed in \cite{Mancuso_2017}, and Paolo Ciliegi and Gianni Zamorani for useful discussion on previous 1.4 GHz source counts. The National
Radio Astronomy Observatory is a facility of the National Science Foundation operated under cooperative agreement by Associated Universities, Inc. H.A., D.vdV. and J.H. acknowledge support of the VIDI research programme with project number 639.042.611, which is (partly) financed by the Netherlands Organisation for Scientific Research (NWO). D.R. acknowledges support from the National Science Foundation under grant number AST-1614213. D.R. also acknowledges support from the Alexander von Humboldt Foundation through a Humboldt Research Fellowship for Experienced Researchers. This research made use of ASTROPY, a community developed core Python package for astronomy (\citealt{astropy:2013, astropy:2018}) hosted at http://www.astropy.org/, matplotlib (\citealt{Hunter_2007}), numpy (\citealt{Walt_2011}), scipy (\citealt{Jones_2001}), and of TOPCAT (\citealt{Taylor_2005}).

\software{ASTROPY \citep{astropy:2013, astropy:2018}, Matplotlib \citep{Hunter_2007}, Numpy \citep{Walt_2011}, Scipy \citep{Jones_2001}, TOPCAT \citep{Taylor_2005}}

\appendix

\section{Postage stamp images}
\label{app:postage}
\begin{figure}[h]
\centering
\includegraphics[width=\columnwidth]{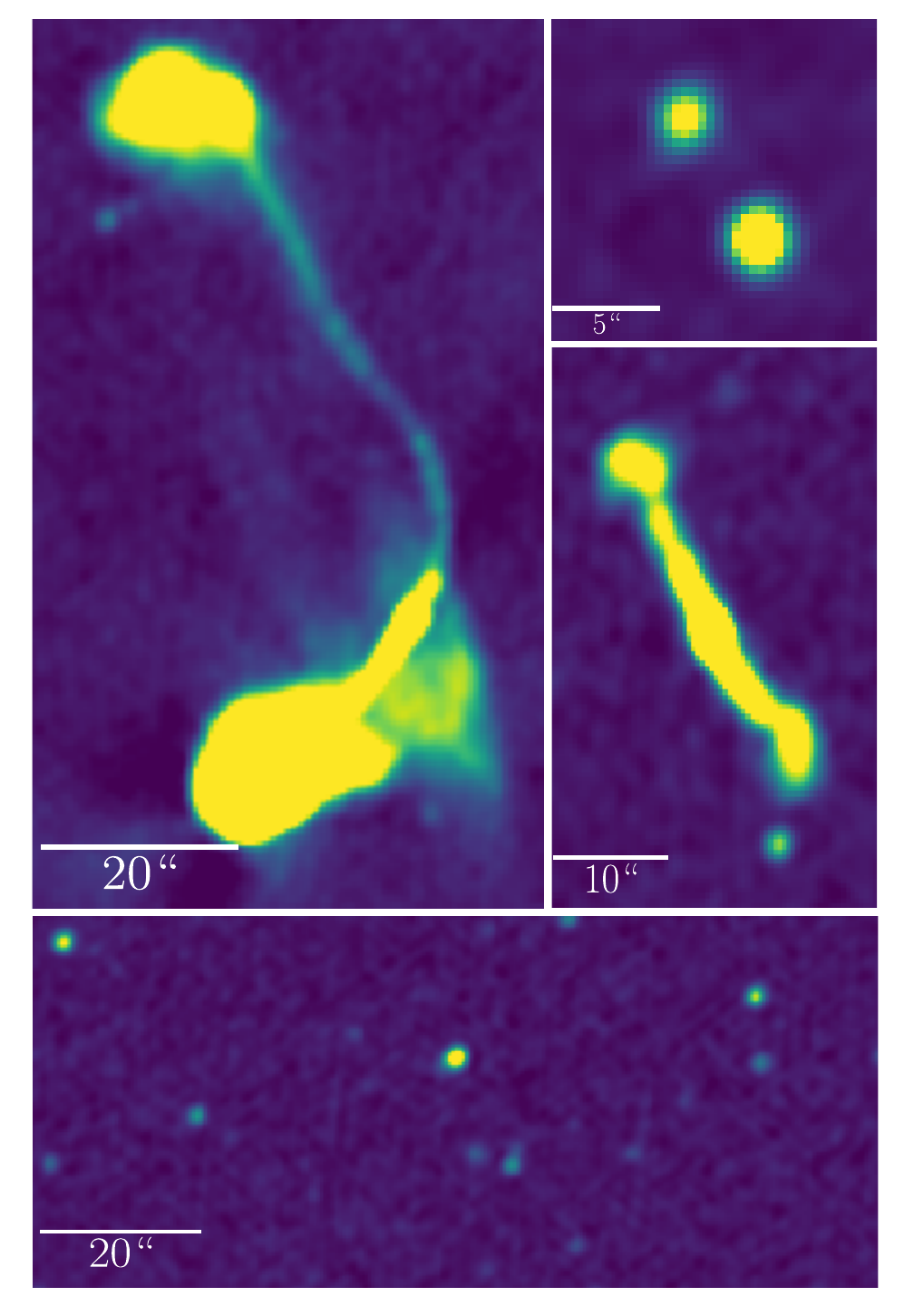}
\caption{Stamps from the $3 \, \rm{GHz}$ continuum map showing examples of extended (left and lower right image), multi-component (upper right) and compact (lower panel) radio sources. The color-scale shows the flux density from $-$3 $\sigma_{\rm local}$ to 50$\sigma_{\rm local}$, where $\sigma_{\rm local}$ is the local r.m.s. noise. }
\label{fig:sources}
\end{figure}

Examples of extended, multi-component and compact sources are shown in Fig. \ref{fig:sources}.

\bibliographystyle{apj}
\bibliography{ref}


\label{lastpage}
\end{document}